\documentclass{aa}
\usepackage{amsmath,amsfonts,amssymb}
\usepackage{textcomp,gensymb}
\usepackage[T1]{fontenc}
\usepackage{aecompl}
\usepackage{ulem}
\usepackage{graphicx}
\usepackage[usenames]{xcolor}
\usepackage{color}
\usepackage{times}
\usepackage{subfig}
\usepackage{stfloats}
\usepackage[utf8]{inputenc}
\usepackage{hyperref}
\usepackage{multirow}
\usepackage{enumerate}
\usepackage{physics}
\usepackage{chemarrow}
\usepackage{tikz}
\usepackage{txfonts}
\usepackage{booktabs}
\usepackage{tablefootnote}
\usepackage{adjustbox}
\usepackage{mathtools}

\usepackage[flushleft]{threeparttable}
\bibpunct{(}{)}{;}{a}{}{,}

\usetikzlibrary{shapes,arrows}
\usetikzlibrary{shapes.geometric}

\newcommand{\rev}[1]{\textcolor{black}{#1}}
\newcommand{\revII}[1]{\textcolor{black}{#1}}

\begin{document} 

    % \title{The Role of an Asymmetric Radiation Field in Shaping Protoplanetary disk Chemistry and Dynamics in Stellar Binary Systems}

    \title{Asymmetric radiation in binary systems: Implications for disk evolution and chemistry}

   % \title{Asymmetric radiation and FUor-like outbursts in stellar binaries: hydrodynamic and temperature profile implications.}

   \subtitle{}

   \author{Pedro P. Poblete\inst{1},
          Nicolás Cuello\inst{1},
          Antoine Alaguero\inst{1},
          Daniel J. Price\inst{2,1},
          Eleonora Bianchi\inst{3},
          Christophe Pinte\inst{1},
          \and
          François Menard\inst{1}.
          }

   \institute{Univ. Grenoble Alpes, CNRS, IPAG, 38000 Grenoble, France;\\
              \email{pedro.poblete@univ-grenoble-alpes.fr};
         \and
             School of Physics and Astronomy, Monash University, Vic. 3800, Australia;
        \and
             INAF, Osservatorio Astrofisico di Arcetri, Largo E. Fermi 5, I-50125, Firenze, Italy.}

   \date{Received September XX, XXXX; accepted March XX, XXXX}

% \abstract{}{}{}{}{} 
% 5 {} token are mandatory
 
  \abstract
   {Current models of binary systems often depend on simplified approximations of the radiation field, which are unlikely to accurately capture the complexities of asymmetric environments. 
   % \cp{\sout{Computational simulations are utilized to model protoplanetary disk environments and to understand their complex architectures. Models of stellar binary systems and their environments often use simple assumptions on their geometry. A proper understanding of the asymmetries in binary systems requires the implementation of a framework beyond these assumptions.} I find this first paragraph too long and convoluted. I much prefer short abstract. Can we not just say: "current models of binary systems rely on simple approximations for the radiation field that are unlikely to be valid in asymmetric environments." ?}
   }
   {We investigate the dynamical and chemical implications of a 3D asymmetric radiation field that accounts for the optical properties of sub-structures present in a protoplanetary disk, as well as the inclusion of a secondary radiation source in binary systems.}
   {We conducted a series of 3D-SPH hydrodynamical simulations using {\texttt Phantom}, coupled with the 3D Monte Carlo radiative transfer code {\texttt Mcfost}, to compute \rev{disc temperatures on-the-fly}. We explored different binary-disk orientations ($0\degree$ and $30\degree$) for an eccentric binary, along with a constant dust-to-gas ratio and dust as a mixture prescription. We also simulated an outburst event as an example of a drastic increase in luminosity.}
   {Heating from the secondary star inflates the outer disk, increasing the aspect ratio facing the companion by about 25\% in inclined configurations compared to 10\% in coplanar ones. Dust settling in the mid-plane enhances extinction along the disk plane, making the coplanar configuration cooler than the inclined one on the side of the disk facing the companion. Additional heating causes a shift in the snow line for species with freeze-out temperatures below 50 K, depending on the disk-binary inclination and binary phase. During outbursts, the aspect ratio doubles on the star-facing side and increases by 50\% on the opposite side in inclined cases. The snow line shift would impact all the species considered in the outburst case.
}
   {Protoplanetary disk heating in binary systems depends on stellar properties, the binary phase, and disk local and global characteristics. This results in temperature asymmetries, especially during secondary star outbursts, leading to variations in aspect ratio and snow lines that can affect chemistry and planet formation.}
{}

% SPH simulations play a crucial role in modeling astrophysical environments such as galaxies, giant molecular clouds, and protoplanetary disks, among others. These simulations solve a set of motion equations in conjunction with an equation of state that links pressure and temperature. Despite a few exceptions, most computational models have assumed a symmetric radiation and temperature field generated by single stars, even in multiple star systems, to determine the hydrodynamics of the surrounding material. In this study, we model the complete asymmetric radiation field of stars using the Monte Carlo radiative transfer code MCFOST and compute the hydrodynamics with the 3D-SPH code PHANTOM. We simulate two configurations to investigate the impact of asymmetric radiation on the circumprimary gas material: a binary system composed of two twin solar-like stars, and two stars with unequal mass, where one experiences an outburst that increases its luminosity by several orders of magnitude. The results reveal highly heterogeneous temperature profiles, with vertical stratification achieved and heating contributions from the disk's outermost regions due to the secondary star resulting in puffed outer vertical areas. These effects have significant implications for the disk's chemistry by shifting the ice line as a function of the binary phase, with the strongest effects observed during an outburst.

   \keywords{Hydrodynamics --- Radiative transfer
 --- Protoplanetary disks --- Methods: numerical
 --- (Stars:) binaries: general  ---  Astrochemistry 
}
   
   \titlerunning{Asymmetric radiation in binary systems}
   \authorrunning{Pedro P. Poblete et al.}
   \maketitle 
%
%-------------------------------------------------------------------\date{Accepted ... Received ...}

% \author[Poblete et al.]{\parbox{\textwidth}{Pedro P. Poblete\thanks{Email: pedro.poblete.rivera@uni-jena.de},
% Tim D. Pearce, 
% Antranik A. Sefilian,
% et al.}
% \vspace{0.15cm}
% \\
% Astrophysikalisches Institut, Friedrich-Schiller-Universität Jena, Schillergäßchen 2–3, 07745 Jena, Germany
% %\\
% }
\label{firstpage}

%%%%%%%%%%%%%%%%%%%%%%%%%%%%%%%%%%%%%%%%%%%%%%%%

\section{Introduction}
\label{sec:intro}

\rev{Protoplanetary disks (PPDs) are circumstellar structures naturally formed during the early stages of star formation and composed of gas and dust. 
%These disks originate from the gravitational collapse of giant molecular clouds, where residual angular momentum causes the formation of a flattened, rotating disk around a central protostar. 
Disks are where planets form: solid dust particles collide and coagulate to form planetesimals, which can then grow through accretion into protoplanets and eventually full-fledged planetary bodies; the physical and chemical processes within the disk directly determine the formation pathways of planets and influence the architecture of the resulting planetary system} \citep[see for example][]{Armitage2018}.

The environment plays an important role in the formation and evolution of PPDs. Star formation frequently occurs within dense clusters, leading to a high multiplicity of stellar systems \citep{Duchene&Kraus2013,Reipurth+2014}. Binary stars are common, accounting for approximately one-third of all stellar systems, although this ratio varies with stellar properties \citep{Raghavan+2010, Tokovinin2014, Tokovinin2021, Offner+2023}. The presence of a second star modifies the environment surrounding a PPD, introducing asymmetries in the gravitational field and radiative heating. These perturbations can affect the structure, dynamics, and evolution of the PPD. 

PPDs in binary stellar systems exhibit diverse structures, broadly categorized into two main configurations.  In circumstellar disks (CSDs), the disk orbits a single star, while in circumbinary disks (CBDs), the disk encircles both stars. The dynamic interplay between the binary and the PPD is sensitive to the binary's properties, including the stellar masses, orbital separation, eccentricity, and inclination.  These parameters influence the evolution and structure of the disk. In circumbinary disks, the effects include, but are not limited to, disk misalignments, the formation of eccentric cavities, the creation of dust over-densities, the generation of disk warps, and even significant disk breaking \citep{Kley+2008a, Kley+2008b, Lodato&Price2010, Nixon+2013, Facchini+2013, Dunhill+2015, Ragusa+2017, Poblete+2019, Hirsh+2020}. Conversely, circumstellar disks in binary systems also experience dynamical perturbations. The gravitational pull from the companion star can shape the CSD via tidal truncation \citep{Paczynski1977}, Lindblad resonances \citep{Lindblad1941}, and von Zeipel-Kozai-Lidov oscillations \citep{Martin2014, Fu+2015, Franchini+2019}. \rev{A review of dust dynamics in disks in stellar binaries, along with a detailed catalog of imaged CSDs and CBDs, can be found in \citet{Zagaria+2023} and \citet{Cuello+2025}, respectively.}

Although progress has been made in modeling the dynamical interactions between binary systems and PPDs using hydrodynamic simulations \citep{Artymowicz&Lubow1994,Larwood1996}, a common simplification is to assume an isothermal disk (e.g. \citealt{Lodato&Price2010,Hirsh+2020}), which neglects radiative heating and cooling processes within the disk. This simplification inaccurately predicts the disk's thermal structure and its influence on other physical processes. \rev{For instance, \citet{Muley+2024} recently studied the observational signatures of embedded planets in disks and \citet{Rowther+2024} provided synthetic observations of a radiative-dominated circumbinary disk with the $\beta$-cooling prescription}.

Accurately representing a disk's temperature profile is critical for several reasons.  Correctly estimating the disk's vertical height relies on this temperature profile \citep{Pringle1981,Kenyon&Hartmann1987, Chiang&Goldraich1997}, as does determining the location of snow lines. Snow lines mark the transition of given molecules from the gaseous state to solid ice. This phase transition affects the local volatile chemical concentration, impacting the building of planetary atmospheres \citep{Oberg+2011,Madhusudhan2012, Cridland+2020} and dust dynamics and growth \citep{Gundlach&Blum+2015, Okuzumi&Tazaki2019}. Snow line positions are determined by the disk's optical, thermal, and chemical properties and the stellar energy flux and are further modified by the presence of multiple stars. \rev{The snow line displacement can be observed during episodic stellar outbursts, which temporarily increase the stellar flux incident on the disk \citep{Cieza+2016, Lee+2019}. Stellar companions are often implicated in outbursting systems \citep{Bonnell&Bastien1992, Zurlo+2024}. \citet{Borchert+2022a, Borchert+2022b} modeled radiation-dominated disks in binary systems during FU Orionis-type outbursts, the most intense kind of episodic accretion events. They showed and characterized changes in the position of the water snow line.}

This work investigates the influence of a secondary heating source on the physical structure of a radiation-dominated CSD. Specifically, it examines how considering the second star's radiation in a binary system affects the disk's vertical structure and temperature profile. This paper is structured as follows: Section \ref{sec:methods} describes the numerical methods and parameters; Section \ref{sec:results} presents the findings; Section \ref{sec:diskussion} offers a discussion of the results. Our conclusions are listed in Section \ref{sec:conclusion}.

\section{Methods}
\label{sec:methods}

We conducted \rev{a total of 14} three-dimensional hydrodynamical simulations of CSDs using the {\texttt Phantom} smoothed particle hydrodynamics (SPH) code \citep{PricePH+2018b}, along with \rev{on-the-fly} radiation field calculations provided by {\texttt Mcfost} \citep{Pinte+2006,Pinte+2009}. We performed two types of simulations: one involving pure gas SPH particles and the other treating dust particles using the dust one-fluid (or dust as a mixture) prescription \citep{Laibe&Price2014b,Laibe&Price2014a,Price&Laibe2015,Ballabio+2018,Hutchison+2018}.

% \cp{I do not think there is a need to re-descibe the codes in so much detail. We do not need the equations (unless you changed something in the codes). I think this section could be much shorter. phantom + mcfost has already been used in several papers. I would refer to those instead.\\
% For the RT part, the choice of equations is also a bit random. You go into very specific details about the opacities and optical depth, but do not indicate the main RT equation.\\
% I think we can simply say mcfost uses a MC method to sample the radiation field and produces observables by ray-tracing the specific intensity}

\subsection{Basic framework of hydrodynamical and radiative computations}
\label{sec:hydro_eqs}

{\texttt Phantom} models the positions and velocities of gas particles using a Lagrangian approach by solving the equations:
\begin{eqnarray}
   \dv{\boldsymbol{v}}{t} &=& -\frac{\nabla P}{\rho} + \Pi_{\rm shock} + \boldsymbol{a}_{\rm ext} + \boldsymbol{a}_{\rm sink},\\
   \dv{u}{t} &=& \frac{P}{\rho^2}\dv{\rho}{t} + \Lambda_{\rm shock},
\end{eqnarray}
where $\boldsymbol{v}$ is velocity, $u$ is specific internal energy, $P$ is pressure, and $\rho$ is density. The accelerations $\boldsymbol{a}_{\rm ext}$ and $\boldsymbol{a}_{\rm sink}$ correspond to external forces and sink particles, respectively. Dissipation terms $\Pi_{\rm shock}$ and $\Lambda_{\rm shock}$ account for shock effects. Accurate velocity and energy variations depend on the equation of state (EOS), typically for an ideal gas:
\begin{equation}
    P = (\gamma -1) \rho u,
\end{equation}
with $\gamma$ as the adiabatic index. In practice, temperature can be related to internal energy from the ideal gas law, giving
\begin{equation}
T_{\rm gas} = \frac{\mu m_H (\gamma - 1)}{k_{\rm B}} u, \label{eq:T_to_u}
\end{equation}
where $\mu = 2.38$ is the assumed mean molecular weight appropriate to a disk consisting mostly of molecular hydrogen, $m_H$ is the atomic mass of Hydrogen, and $k_{\rm B}$ is the Boltzmann constant. Assuming an isothermal distribution simplifies the temperature and internal energy profiles so they only depend on radius, centered on the primary heating source. However, this neglects the disk’s optical, thermal, and chemical properties, which is problematic in systems with multiple heating sources, such as binaries. 

{\texttt Mcfost} is a 3D Monte Carlo radiative transfer code that divides the space into Voronoi cells centered on SPH particles, with physical properties constant within each cell. Radiation is modelled via photon packets originating from star and dust thermal emission, which interact with the medium through scattering, absorption, or re-emission. Dust optical properties are derived from Mie theory, assuming spherical, homogeneous grains, with opacities computed by integrating over grain size distributions. Under local thermodynamic equilibrium (LTE), the cell’s temperature $T_i$ is derived via the mean intensity approach \citep{Lucy1999}:
\begin{equation}\label{eq:rad_eq}
    \int_0^\infty \kappa_i^{\rm abs}(\lambda) B_\lambda(T_i){\rm d}\lambda = \frac{L_{\rm star}}{4 \pi V_i N_\gamma}\sum_{\lambda, \gamma} \kappa_i^{\rm abs}(\lambda)\Delta l_\gamma + \Lambda_{\rm heat},
\end{equation}
where $\kappa_i^{\rm abs}$ is the cell's opacity, $B_\lambda$ is the blackbody intensity, $L_\mathrm{star}$ the stellar luminosity, $V_i$ the cell volume, $N_\gamma$ photon packets, $\Delta l_\gamma$ is the distance over which a single photon travels before being extinguished, and $\Lambda_{\rm heat}$ additional heating mechanisms. This process allows {\texttt Mcfost} to iteratively compute the dust temperature distribution accounting for complex interactions within the disk. We then assumed the dust temperature is equal to the gas temperature in order to update the internal energy of the gas in {\texttt Phantom} by inverting Eq.~(\ref{eq:T_to_u}).

\subsection{Simulation setup}
\label{sec:Phantom_setup}

\begin{table*}
\centering
\begin{adjustbox}{}

\begin{threeparttable}
\caption{Simulation setups.}
% \begin{tabular}{ccccccccccc}
\begin{tabular*}{\textwidth}{@{\extracolsep{\fill}}ccccccccccc@{}} 

\hline
\hline
% \noalign{\hrule height 1.5pt}
\multicolumn{11}{c}{}\\
\multirow{2}{*}{Simulation} & \multirow{2}{*}{Name} & Radiative disk & \multirow{2}{*}{Dust prescription} & Outer & $N_{\rm sph}$ & $N_{\gamma}$ & $M_{\rm star}$ & $T_{\rm eff, star}$ & $R_{\rm star}$ & $L_{\rm star}$ \\
& & {(coupled \texttt Mcfost)} & & taper  & $[10^6]$ & $[10^9]$ & $[M_{\odot}]$ & $\rm [K]$  & $[R_{\odot}]$ & $[L_{\odot}]$ \\

\hline

\multirow{2}{*}{$1$} & \multirow{2}{*}{Iso-Bin-q1} & \multirow{2}{*}{No} & \multirow{2}{*}{Dust-free} & \multirow{2}{*}{Yes}   & \multirow{2}{*}{0.5} & \multirow{2}{*}{-} & 1.0 & - & - & - \\
  &  &   &  &   &  &   &  1.0 & - & - & - \\

\hline

\multirow{2}{*}{$2$} & \multirow{2}{*}{OS-Bin-q1} & \multirow{2}{*}{Yes} & \multirow{2}{*}{Dust-free} & \multirow{2}{*}{Yes}   & \multirow{2}{*}{0.5} & \multirow{2}{*}{0.7} & 1.0 & 5800 & 1.0 & 1.0 \\
  &  &   &  &   &  &   & 1.0 & - & - & - \\

\hline

\multirow{2}{*}{$3$} & \multirow{2}{*}{Bin-q1} & \multirow{2}{*}{Yes} & \multirow{2}{*}{Dust-free} & \multirow{2}{*}{Yes}   & \multirow{2}{*}{0.5} & \multirow{2}{*}{0.7} & 1.0 & 5800 & 1.0 & 1.0 \\
  &  &   &  &   &  &  & 1.0 & 5800 & 1.0 & 1.0 \\

\hline

\multirow{2}{*}{$4$} & \multirow{2}{*}{NT-Bin-q1} & \multirow{2}{*}{Yes} & \multirow{2}{*}{Dust-free} & \multirow{2}{*}{No}   & \multirow{2}{*}{0.5} & \multirow{2}{*}{0.7} & 1.0 & 5800 & 1.0 & 1.0 \\
  &  &   &  &   &  &  & 1.0 & 5800 & 1.0 & 1.0 \\

\hline

\multirow{2}{*}{$5$} & \multirow{2}{*}{Dust-Bin-q1} & \multirow{2}{*}{Yes} & \multirow{2}{*}{One-fluid} & \multirow{2}{*}{No}   & \multirow{2}{*}{0.5} & \multirow{2}{*}{2.0} & 1.0 & 5800 & 1.0 & 1.0 \\
  &  &   &  &   &  &  & 1.0 & 5800 & 1.0 & 1.0 \\

\hline

\multirow{2}{*}{$6$} & \multirow{2}{*}{Bin-q05} & \multirow{2}{*}{Yes} & \multirow{2}{*}{Dust-free} & \multirow{2}{*}{Yes}   & \multirow{2}{*}{1.0} & \multirow{2}{*}{1.0} & 1.0 & 5800 & 1.0 & 1.0 \\
  &  &   &  &   &  &   & 0.5 & 3700 & 0.46 & 0.036 \\

\hline

\multirow{2}{*}{$7$} & \multirow{2}{*}{FU-Bin-q05} & \multirow{2}{*}{Yes} & \multirow{2}{*}{Dust-free} & \multirow{2}{*}{Yes}   & \multirow{2}{*}{1.0} & \multirow{2}{*}{4.0/2.0\tnote{a}} & 1.0 & 5800 & 1.0 & 1.0 \\
  &  &   &  &   &  &  & 0.5 & 3700 & 13.5 & 30.0 \\

\hline
\hline
\end{tabular*}
\begin{tablenotes}
\small
   \item \textbf{Notes.} \rev{Each simulation is tagged with a unique name that encodes key configuration details. The used tags are: Iso (isothermal temperature profile), OS (one-source irradiation), NT (no outer taper), and FU (FU Orionis-like outburst). It is also written whether the disk features an initial exponential outer taper. The parameter $N_{\rm sph}$ denotes the number of SPH particles, while $N_\gamma$ represents the total number of photon packets used in the radiative transfer calculations. The stars have a mass $M_{\rm star}$ and, in simulations where {\texttt{Mcfost}} is enabled, are further characterized by an effective temperature $T_{\rm eff, star}$, luminosity $L_{\rm star}$, and radius $R_{\rm star}$. All simulations are carried out in both coplanar ($0^\circ$) and inclined ($30^\circ$) binary-disk configurations, yielding a total of 14 simulations.
   \item[a] For the coplanar/inclined case, respectively.
   }     
\end{tablenotes}
\label{tab:bin_params}
\end{threeparttable}
\end{adjustbox}

\end{table*}

\begin{table}
\centering

\begin{threeparttable}
\caption{Disk parameters utilized in constructing the simulation.}
\begin{tabular*}{0.485\textwidth}{@{\extracolsep{\fill}}lcclcc@{}} 
\hline
\hline
% \noalign{\hrule height 1.5pt}
\multicolumn{6}{c}{}\\

\multicolumn{3}{c}{Gaseous disk } &
\multicolumn{3}{c}{Dusty disk }\\
\cmidrule(lr){1-3}
\cmidrule(lr){4-6}

Name & Unit & Value & Name & Unit & Value \\
\cmidrule(lr){1-3}
\cmidrule(lr){4-6}
$\Sigma_0$ & g cm$^{-2}$ & $10^2$ & $\Sigma_{0,\rm dust}$ & g cm$^{-2}$ & $1.0$ \\
$T_0$ & K & $127.9$ & $\rho_s$ & g cm$^{-3}$ & $3.0$ \\
$r_{\rm min}$ & au & 1.25 & $s_{1}$ & mm & $0.2$ \\
$r_{\rm max}$ & au & 12.5 & $s_{2}$ & mm & $1.0$ \\
$r_0$ & au & 5.0 & $s_3$ & mm & $4.6$ \\
\revII{$p_1$} & - & 1.0 & ${\rm St_{1}}$ & $10^{-2}$ & 0.2\\
\revII{$p_2$} & - & 0.5 & ${\rm St_{2}}$ & $10^{-2}$ & 1.1\\
 & & & ${\rm St_{3}}$ & $10^{-2}$ & 5.1\\
\hline
\hline

\end{tabular*}

\begin{tablenotes}
\small
    \item \textbf{Notes.} \textit{Gaseous disk}: we initialised the gas disk with a surface density $\Sigma_0$ and temperature $T_0$ at the reference radius $r_0$ by following power laws with exponents \revII{$p_1$ and $p_2$}, respectively (see Equations \ref{eq:sigma_prof} and \ref{eq:T_prof}). The disk extends from $r_{\rm min}$ to $r_{\rm max}$.  \textit{Dusty disk}: The dust disk is initialized equal to the gaseous disk but with a different surface density $\Sigma_{0,\rm dust}$ at $r_0$. The dust disk is composed of three dust grains: $s_1$, $s_2$, and $s_3$ \rev{with the corresponding Stokes number ${\rm St_{1}}$, ${\rm St_{2}}$, and ${\rm St_{3}}$, respectively,} having all of them an intrinsic grain density of $\rho_s$.
\end{tablenotes}

\label{tab:disk_params}
\end{threeparttable}
\end{table}

% Tables \ref{tab:bin_params} and \ref{tab:disk_params} summarize the stellar and disk parameters along with the resolution of the simulations.

% \subsubsection{Initial conditions and simulation parameters}
% \label{sec:initial_conditions}

We conducted \rev{14} simulations for different stellar parameters, binary-disk inclinations, and disk-thermal prescriptions. The simulation architecture consists of a stellar binary system where the primary star hosts a circumstellar disk. We modelled the two stars as sink particles \citep{Bate+1995}, with accretion radii set to 1 au for both. \rev{All binary systems started with a semi-major axis of 50 au and an eccentricity of 0.5. For each configuration, we explored two inclination angles relative to the circumstellar disk: a coplanar ($0\degree$) and an inclined case ($30\degree$). We considered two binary mass ratios ($q$): one for an equal-mass binary system ($q=1$) composed of twin Sun-like stars, and another with $q=0.5$, which involves a primary Sun-like star and a secondary star with half its mass. We used standard Sun-like stellar parameters and assumed main-sequence stars to derive their effective temperatures and radii for all 1.0 $M_\odot$ stars. For all 0.5 $M_\odot$ secondary stars, we similarly assumed main-sequence characteristics consistent with red dwarf stars \citep{Boyajian+2012, Baraffe+2015}. The simulation Bin-q05 represents a binary system where the secondary star remains in a quiescent state throughout the run, whereas FU-Bin-q05 models an outbursting secondary. The latter setup was constructed by only modifying the stellar radiative properties of the companion at the end of the Bin-q05 simulation, thereby preserving the same evolutionary and dynamical configurations of both stars and the disk. In order to create the outbursting state (luminosity increase), we artificially modified the stellar radius ($R_{\rm star}$)} of the secondary star while maintaining a constant effective temperature. The change in radius will only affect the radiative calculations in {\texttt Mcfost} in two aspects: (i) it will increase the bolometric stellar luminosity up to $30\ L_\odot$ without altering the stellar spectrum, as the effective temperature remains constant, and (ii) the starting point of the photon packets will now be set at a new distance of $R_{\rm star}$. \rev{Table~\ref{tab:bin_params} summarizes all the parameters considered in our simulations.}

The outburst events are associated with an increase in the star's accretion rate \citep[see the review][]{Fischer+2023} and are known to have diverse origins \citep{Vorobyov+2021}. We assumed that the most likely point for the companion to trigger an outburst is at pericentre as \citet{Bonnell&Bastien1992} suggest, where it experiences a rise in its accretion rate \citep[see also figure 5 of][]{Poblete+2020}. Therefore, we force the luminosity change to occur at pericentre by the end of the evolution of Bin-q05. As we are mainly interested in the immediate effect of the outburst on the disk, we model the high luminosity phase during a couple of decades only.

% \pp{To be discussed}
% \cp{The method seems a bit random here. Why not using the accretion luminosity ? 
% The outburst will also likely change the spectral type of the emission, so I am not sure how realistic that is. \\
% Also, the choice of a 5700K and 1Rsun for a 1 solar mass is a bit strange
% \\
% I find this even more strange. We should really just use the accretion luminosity. I am a bit confused about what we can learn by just randomly and instantly increasing the radiation field by a factor 1000} 

\revII{The gas disk is initialized with $N_{\rm sph}$ SPH particles in Keplerian rotation and is irradiated by $N_\gamma$ photon packets emitted from both stars, with each emission occurring at intervals of $\Delta t_\gamma$, corresponding to the timestep between {\texttt Mcfost} calls} (see Table \ref{tab:bin_params}). The total number of photon packets is distributed between the two stars weighted by their luminosities. The explanation for the heterogeneous $N_\gamma$ values will be discussed further in Section \ref{sec:ph+mc}. The initial gas surface density and temperature profiles follow
\begin{eqnarray}
    \Sigma(r) &=& \Sigma_0\ \left(\frac{r}{r_0}\right)^{-\revII{p_1}}\left(1-\sqrt{\frac{r_{\rm min}}{r}}\right) \underbrace{\left(1-\exp{r-r_{\rm max}}\right)}_\text{\clap{\rev{outer taper}}} , \label{eq:sigma_prof}\\
    T(r) &=& T_0\ \left(\frac{r}{r_0}\right)^{-2\revII{p_2}},\label{eq:T_prof}
\end{eqnarray}
using the values in Table \ref{tab:disk_params}. \rev{Exponentially tapering the outer edge of the disc in NT-Bin-q1 and Dusts-Bin-q1 results in negative dust-to-gas ratios. Therefore, the discs in NT-Bin-q1 and Dusts-Bin-q1 are not exponentially tapered at the outer edge. We note that, by omitting the taper and using a fixed $\Sigma_0$ at $r_0$ and disk radius, these disks are roughly 13\% more massive than the other simulations.} We initially set the disk's temperature profile to be vertically isothermal for all simulations, following a power law that results in an aspect ratio of $H/r = 0.05$ at any radius. Also, we set the SPH viscosity parameter $\alpha_{\rm AV} = 0.13$, which corresponds to a mean Shakura-Sunyaev disk viscosity \citep{Shakura&Sunyaev73} of $\alpha_{\rm SS} \approx 5 \times 10^{-3}$ \citep[cf.][]{Lodato&Price2010}. 

We explore three thermal-disk prescriptions:
\begin{itemize}
    \item \text{Radially isothermal}: the temperature only depends on the radial distance following Equation \ref{eq:T_prof}.  
    \item \text{Radiative dust-free}: The temperature is set by {\texttt Mcfost} assuming a constant gas-to-dust ratio of 100 in the gas density profile provided by {\texttt Phantom}. We use the Mie theory for dust grains and we consider 100 different grain species that follow a grain-size distribution by following a power law ${\rm d}n/{\rm d}s\propto s^{-3.5}$ between $0.03\ \micro$m and $1$ mm. 
    \item \text{Radiative one-fluid dust}: The temperature is set by {\texttt Mcfost} with dust distribution given by {\texttt Phantom}. The dust disk is initialized equal to the gas disk. We consider three dust species $s_1$, $s_2$, and $s_3$, in the limit of dust diffusion as presented in \citet{Hutchison+2018} and \citet{Ballabio+2018}. See \textit{Dusty disk} in Table \ref{tab:disk_params}.
\end{itemize}

In the radiative dust-free and one-fluid dust prescriptions, the dust composition is assumed to be mainly silicates. We evolve the simulations until they reach 10 binary orbits, corresponding to $\approx2500$ years for Bin-q1 and $\approx2700$ years for Bin-q05. Finally, we set the frequency for updating temperatures in {\texttt Phantom} to 100 times per binary orbit \rev{during the whole simulation. This corresponds to an {\texttt Mcfost} call} every $\Delta t_\gamma =\{ 2.5, 2.8\}$ years for Bin-q1 and Bin-q05, respectively. 
% \begin{equation}
%     \omega = \frac{\sqrt{GMa(1-e^2)}}{a^2(1-e^2)^2} \cdot (1+e\cos(\nu))^2
% \end{equation}

% \subsection{Radiative transfer calculations setup}
% \label{sec:Mcfost_setup}

% We assume stars emit as a blackbody, and Table \ref{tab:sim_params} lists the star radiative properties. Due to our simulation not considering dust particles, we construct the dust population using the SPH gas particle distribution for only radiative proposes. We set the gas-to-dust ratio to 100, and we consider 100 different grain species that follow a grain-size distribution by following a power law ${\rm d}n/{\rm d}s\propto s^{-m}$ between $s_{\rm min}=0.03\ \mu$m and $s_{\rm max}=1\ $mm with $m=3.5$.
% The Monte Carlo radiative transfer is computed by considering a number of photons package equals $7\times10^8$. 

% Define block styles
\tikzstyle{block} = [rectangle, draw, node distance=2.0cm,
    text width=7em, text centered, rounded corners, minimum height=3em]
\tikzstyle{line} = [draw, -latex']
\tikzstyle{cloud} = [draw, ellipse, node distance=1cm,
    minimum height=1cm, text width=8.5em, text badly centered]

\tikzstyle{dia} = [draw, diamond, inner sep=-3pt, node distance=2.0cm, text width=8.5em, text badly centered, aspect=2]
    
\tikzstyle{block2} = [rectangle, draw, node distance=5cm,
    minimum height=0.25cm, text width=8em, text badly centered]
    
\tikzstyle{block3} = [rectangle, draw, node distance=2.0cm,
    text width=8em, text badly centered, minimum height=0.25cm]

 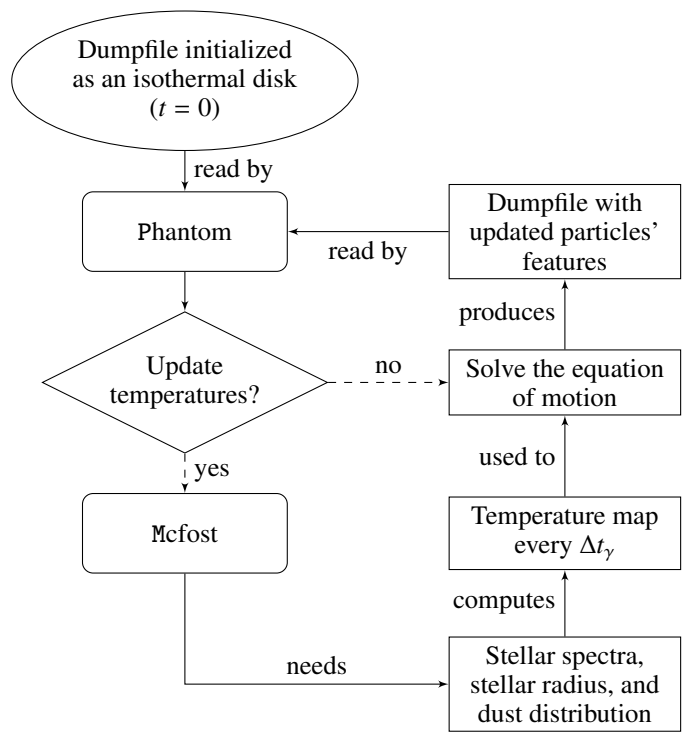
\begin{figure}
    \centering
    \begin{tikzpicture}[node distance = 0cm, auto]
        % Place nodes
        \node [cloud] (init) {\rev{Dumpfile} initialized as an isothermal disk\\($t=0$)};      \node [block, below of=init] (PH) {{\texttt Phantom}};
        \node [dia, below of=PH] (cond) {Update temperatures?};
        \node [block, below of=cond] (MC) {{\texttt Mcfost}};
        \node [block2, right of=MC] (Tmap) {Temperature map\\ \rev{every $\Delta t_\gamma$}};
        \node [block3, below of=Tmap] (Spec) {Stellar spectra, stellar radius, and dust distribution};
        \node [block2, right of=PH] (snap) {\rev{Dumpfile with updated particles' features}};
        \node [block3, below of=snap] (eq) {Solve the equation of motion};
        % Draw edges
        %\path [line] (casos) -| node [near start] {yes} (update);
        %\path [line] (update) |- (n);
        %\path [line] (casos) -- node {no}(stop);
    
        \path [line] (init) -- node {read by}(PH);
        \path [line] (Spec) -- node {computes}(Tmap);
        \path [line] (Tmap) -- node {used to}(eq);
        \path [line] (PH) -- node {}(cond);
        \path [line] (eq) -- node {produces}(snap);
        \path [line] (snap) -- node {read by}(PH);
        \path [line] (MC) |- node [near end]{needs}(Spec);
        \path [line,dashed] (cond) -- node {yes}(MC);
        \path [line,dashed] (cond) -- node {no}(eq);
    \end{tikzpicture}
    \caption{Flowchart illustrating the coupling between the SPH code {\texttt Phantom} and the radiative transfer code {\texttt Mcfost}. An ellipse represents the initial process, the main processes are shown in rectangles, and the decision node is depicted as a diamond.}
    \label{fig:DFlux}
 \end{figure}

\subsection{Coupling and limitations between {\texttt Phantom} and {\texttt Mcfost}}
\label{sec:ph+mc}

% \cp{this mostly duplicates the mcfost paper in prep. I think we should summarize here instead and refer to the paper. It would also be nice to cite the papers which already used phantom+mcfost : Bec's, Ellie's, Sahl's and mine on HD97. Bec 2019 paper and Ellie's paper in particular explore similar topics as your paper so they definitely need to be cited}

\rev{{\texttt Phantom} computes the evolution of the hydrodynamic variables in our simulations, while {\texttt Mcfost} calculates the whole radiation field and resulting temperature profiles. In our simulations, {\texttt Phantom} initialized an isothermal vertical profile and provided the positions of the sink and SPH particles to {\texttt Mcfost}. Depending on the dust prescription, the dust content in each Voronoi cell was either set via a fixed dust-to-gas ratio value or derived from {\texttt Phantom} outputs. {\texttt Mcfost} required a stellar spectrum and radius to define the photon packets' properties. These photon packets were propagated in the disk to get the temperature structure, which was then returned to {\texttt Phantom} to update hydrodynamics} \citep[explained in detail in][]{Pinte+2019,Nealon+2020a}, as summarized in Figure \ref{fig:DFlux}. Cells evolved adiabatically between {\texttt Mcfost} calls \rev{(i.e., every $\Delta t_\gamma$)}, allowing for adiabatic work.
% \pp{This is not the right way to account for shock heating as the temperature is continually over-ridden by the calls to MCFOST. In recent code we include the shock heating as a source term in MCFOST. Please check if this is what you meant, or something else?}
Unlike \citep{Borchert+2022a,Borchert+2022b}, we did not model accretion luminosity, so temperature changes depended solely on stellar position, luminosity, adiabatic work, and disk properties.

Our modelling required frequent radiative transfer calls to compute hydrodynamics accurately, ensuring (i) that the dynamical time exceeds the interval between temperature updates \citep[see][]{Rowther+2024}, and (ii) that temperature shifts caused by the secondary star’s movement relative to the disk’s outer edge are minimized. The chosen timestep $\Delta t_\gamma$ allowed particles at $\approx2$ au to complete one orbit around the primary star before their temperatures were recalculated \rev{by {\texttt Mcfost}}; meanwhile, at $r_{\rm max}$, the temperature was updated 25 times along a local particle's orbit.

Finally, insufficient photons can lead to artifacts due to non-converging cell temperatures. This issue presents itself in two ways: (i) cells without any energy heating default to a temperature of 1 K, and (ii) cell temperatures fail to converge. Both factors risk causing unphysical collapse and fragmentation. Dense regions where photons cannot penetrate are especially prone. To prevent this, we utilize a large, variable number of photons and monitor the simulation, increasing photon packets in dense mid-plane areas as necessary. For example, we change the value $N_\gamma/N_{\rm sph} = 1400$ in \rev{NT-Bin-q1 up to a ratio of 4000 in Dust-Bin-q1 to ensure proper irradiation in the one-fluid simulation} while keeping computational costs manageable.

\begin{figure*}
\centering
\begin{center}
    \includegraphics[width=1.0\textwidth]{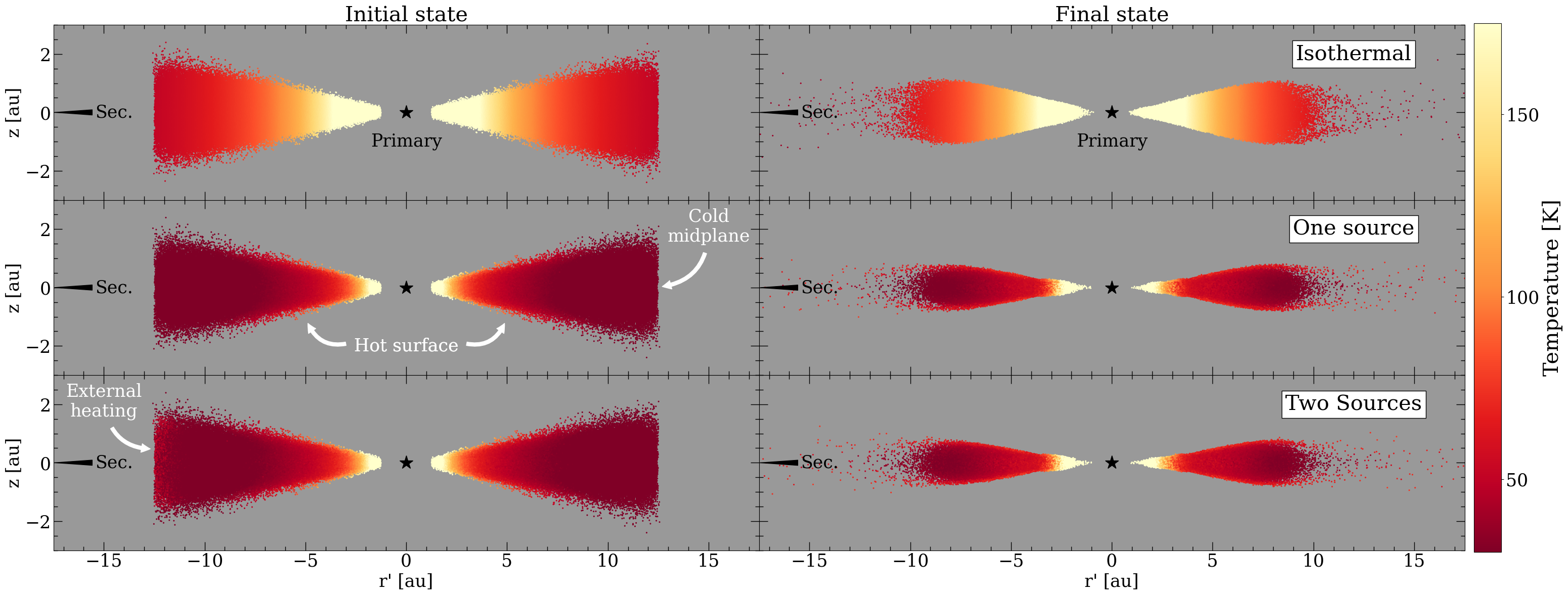} 
    \caption{\rev{Temperature distributions of SPH particles (in the \texttt{Phantom} dumpfile) in the CSD are shown in the $r'$–$z$ plane for three heating scenarios: (top) vertically isothermal; (middle) irradiation from the primary star only; and (bottom) irradiation from both stars. These correspond to the Iso-Bin-q1, OS-Bin-q1, and Bin-q1 simulations, respectively. The left and right columns show the simulation's initial and final states. A black star marks the primary star's position, while a black arrow points to the secondary star.}}
    \label{fig:Tmap_coplanar}
\end{center}
\end{figure*}
\begin{figure*}
\centering
\begin{center}
    \includegraphics[width=1.0\textwidth]{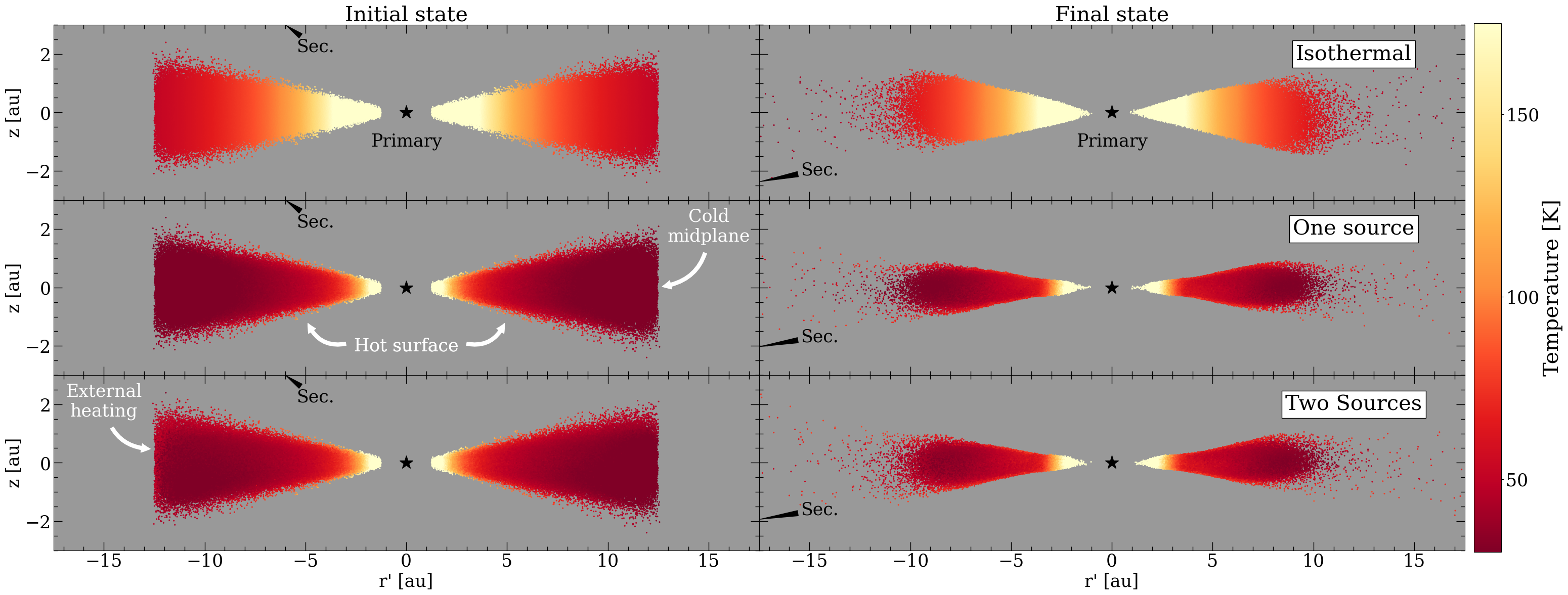} 
    \caption{Same as Fig. \ref{fig:Tmap_coplanar} but for a binary-disk orbital inclination of $30\degree$.}
    \label{fig:Tmap_inc}
\end{center}
\end{figure*}

\section{Results}
\label{sec:results}

To clearly visualize the asymmetric heating effects of the secondary star on the circumprimary disk, our simulations use a re-oriented coordinate system. The disk's angular momentum vector is aligned perpendicular to the $xy$-plane, and the secondary star's orbit is constrained to a co-rotational frame; the companion's movement is confined to the $x$ axis and the plane $y=0$ for the coplanar and inclined configuration, respectively. A new radial coordinate, $r'$, is defined based on the standard cylindrical radial coordinate, $r$, to facilitate analysis within this re-oriented framework. The new radial coordinate is defined as a function of the distance between a given particle within the disk and the secondary star's position as follows
\begin{align}
   r'(r) 
    &= \left\{\begin{array}{rl}
             -r, & \text{if it is located within the half of the disk nearest}\\
             &\text{ to the star,}\\
             % &  \\
             r, & \text{if it is located within the half of the disk nearest}\\
             &\text{ to the star.}\\
           \end{array}\right.
\end{align}
Section \ref{sec:iso-rad_comp} compares results from \rev{Iso-Bin-q1, OS-Bin-q1, and Bin-q1 simulations. Section \ref{sec:dust_rad_comp} compares the dust prescriptions represented by NT-Bin-q1 and Dust-Bin-q1 simulations. Finally, the impact of an outburst from the secondary on the CSD is explored in Section \ref{sec:Fuori} based on the Bin-q05 and FU-Bin-q05 simulations. For each of the previous cases, we also investigated the effect of the two misaligned binary-disk configurations ($0\degree$ and $30\degree$).}

\subsection{Isothermal vs radiative-dominated disk}
\label{sec:iso-rad_comp}

\subsubsection{Radial and vertical profiles}

\begin{figure*}
\centering
\begin{center}
    \includegraphics[width=1.0\textwidth]{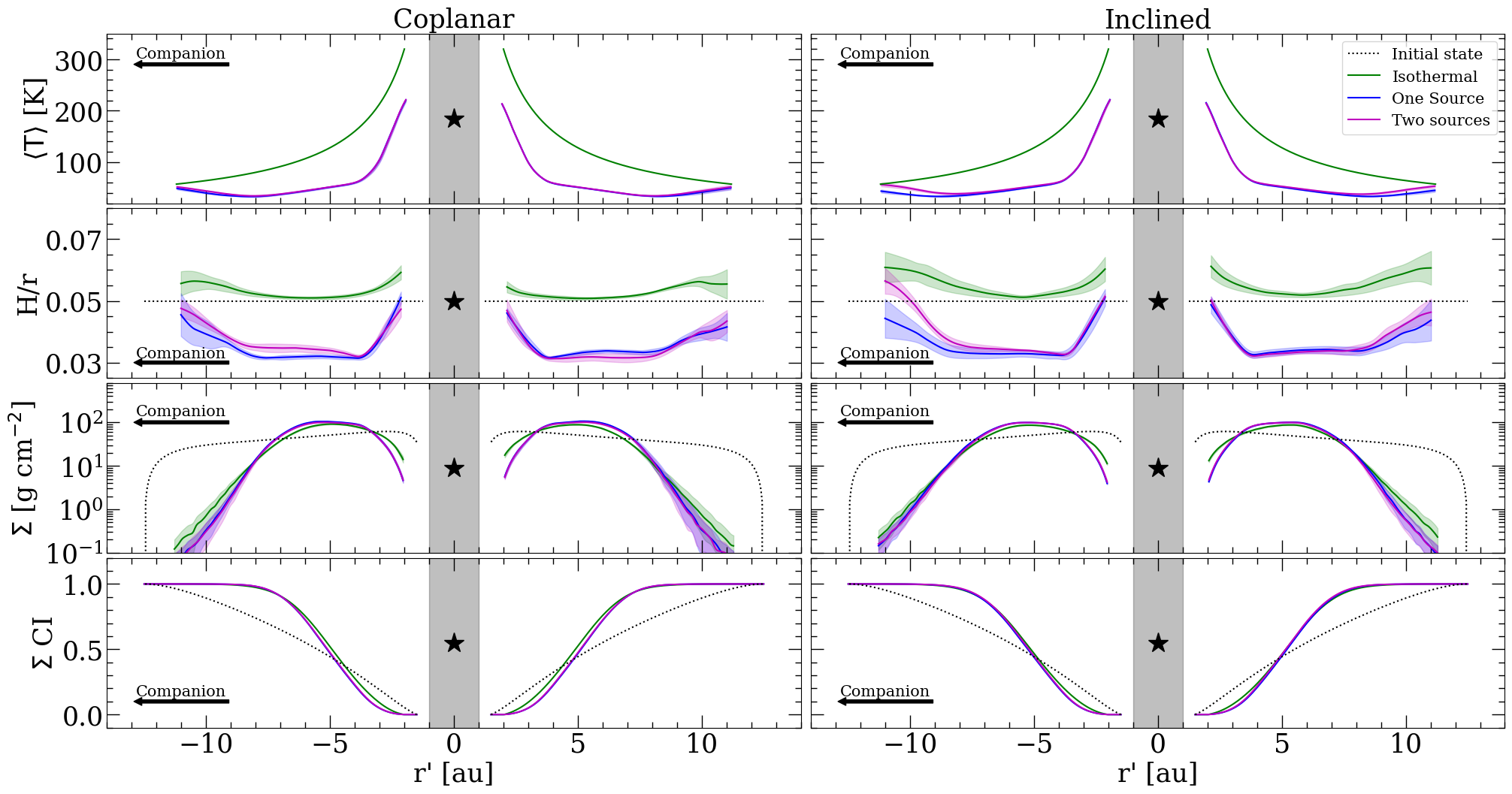} 
    \caption{Figure shows the median radial profile of disk density-weighted average temperatures (first row), aspect ratio ($H/r$, second row), surface density ($\Sigma$, third row), and the normalized surface density cumulative integral ($\Sigma$ CI, fourth row) for a coplanar (left) and inclined by $30\degree$ (right). \rev{Each panel compares simulations with a vertically isothermal temperature profile (Iso-Bin-q1 in green), a one-star heating model (OS-Bin-q1 in blue), and a two-star heating model (Bin-q1 in magenta)}. Solid lines represent the median values along one binary orbit (between the ninth and tenth), while the shaded colour regions illustrate the median absolute deviation. The shaded gray areas denote the primary star's accretion radius. The black arrow indicates the location of the companion star. The dotted line in the second, third, and fourth rows depicts the initial values for the $H/r$, surface density profiles, and the normalized surface density cumulative integral, respectively.}
    \label{fig:hr_sigma_prof}
\end{center}
\end{figure*}

Figures \ref{fig:Tmap_coplanar} and \ref{fig:Tmap_inc} offer a visual comparison of temperature distributions within the CSD, illustrating results from different heating scenarios for coplanar and inclined binary configurations, respectively. In both figures, the radiative simulations with {\texttt Mcfost} \rev{(OS-Bin-q1 and Bin-q1)} reveal vertical thermal stratification caused by density-dependent extinction, with denser mid-plane regions being cooler than surface layers. This behaviour contrasts with the isothermal model \rev{(Iso-Bin-q1)}, which does not incorporate the density-temperature coupling and therefore lacks such vertical temperature variation. Figure \ref{fig:hr_sigma_prof} illustrates the differences in temperature, aspect ratio, and surface density profiles when radiative temperature calculations are applied to both coplanar and inclined binary systems. As in the temperature maps, they reveal significant differences between the radiative and isothermal models. We calculated the scale height ($H$) by fitting Gaussian profiles to the vertical density structure at different disk radii, such as $\rho(z)\propto\exp(-z^2/2H^2)$. 

\rev{Including radiation from the secondary star causes asymmetries in the disk’s temperature distribution in the outer regions compared to the OS-Bin-q1 simulation, where only the primary star heats the disk. In the coplanar setup, the side nearest to the companion experiences an approximate 4 K temperature increase, while the farthest side rises by 2 K, both representing about 10\% increases over the one-source case. In the inclined setup, these differences become more pronounced: the closest side heats up by as much as 12 K, and the farthest side by 6 K, corresponding to relative increases of nearly 35\% and 20\%, respectively. These findings indicate that in inclined systems, the secondary star's radiation more effectively penetrates both the inner and outer regions of the disk. Conversely, in the coplanar scenario, the dense mid-plane greatly absorbs the secondary's radiation, limiting its thermal influence on the disk.}

Radiative disks exhibit a reduced aspect ratio compared to isothermal disks, a direct consequence of their adjustment to maintain hydrostatic equilibrium. This is evident in the decrease from an initial $H/r$ value of 0.05 to approximately 0.03 between 4 and 8 au. In contrast, the isothermal model, by construction, is already in hydrostatic equilibrium, precluding any such later adjustment in aspect ratio. In the isothermal model, changes in the vertical structure are confined to the inner and outer disk rims, where interactions with stars are most pronounced, leading to a decrease in surface density. As we derived the aspect ratio from the vertical density profile, it is dependent on temperature and density. Consequently, these rims exhibit increased $H/r$ values due to constant temperature, as less material supports the initial vertical profile, resulting in local thickening.
% While the radiative models also exhibit vertical variations, these variations are more pronounced and stand in greater contrast to the rest of the disk structure compared to the isothermal model, underscoring the different physical mechanisms at play in each model.\cp{This needs to be quantitative. The current sentence says nothing}

The binary's inclination impacts the disk's vertical structure. In coplanar configurations, the secondary star's radiation primarily heats the near side of the disk, resulting in only a $\sim$10\% increase in the $H/r$ value in this region when comparing one-source and two-source heating models. The far side shows negligible change. Conversely, the rise in the $H/r$ value is considerably larger in inclined systems: approximately 25\% on the near side and 10\% on the far side.  Furthermore, the ability of the secondary star's radiation to penetrate and heat the far side of the disk more effectively in inclined systems leads to an increase in contrast between the borders and the middle disk vertical structure. These differences highlight the critical role of the secondary star's radiation in shaping the vertical structure of the disk, particularly in inclined binary-disk systems, \rev{as discussed above for the temperature.}  

Finally, radiative disks have regions that are denser than those of their isothermal counterparts. These disks show a density concentration toward the central regions, with densities decreasing in both the inner and outer rims. In the simulations, approximately 90\% of the circumprimary disk mass is contained within a radius of about 7 au in both coplanar and inclined cases. However, the isothermal simulations contain 45\% more material within 3 au than the radiative simulations, which reduces to 15\% at 4 au, with both models enclosing the same amount of material by 7 au. While this might initially suggest a dominant role for radiative pressure in shaping this radial profile, especially from the secondary star, the one-source and two-source simulations exhibit this same compact radial profile, discarding such a role. This indicates that radiative pressure effects do not primarily drive the density profile but rather the different temperature profiles inherent to radiative models. 
% This cooling in the inner regions reduces thermal pressure, resulting in the disk's radial extent contraction. Therefore, the compact radial profile of radiative disks arises primarily from the interplay between temperature and pressure gradients induced by the radiative transfer calculations incorporated in the model rather than solely from the effects of radiative pressure.\cp{I do not understand. Why is this simply not do to the change in the temperature (and hence pressure) profile due to the illumination by the primary ? ie T(r) is very different from the one assumed in the isothermal case}

\begin{figure*}
\centering
\begin{center}
    \includegraphics[width=1.0\textwidth]{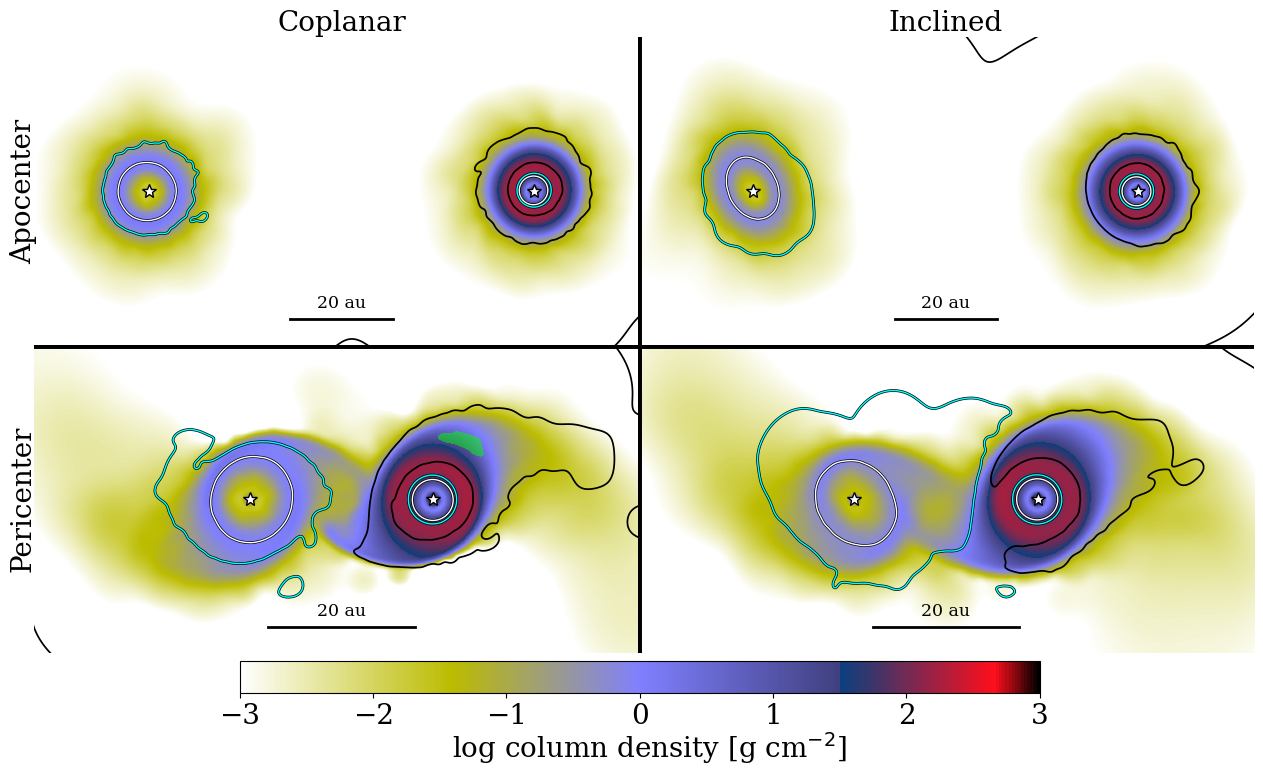} 
    \caption{Column density maps of circumstellar disks in coplanar (left) and inclined (right) binary systems at the apocentre (top) and pericentre (bottom), viewed face-on with respect to the circumprimary disk in Bin-q1. The evolutionary time at the pericentre and apocentre corresponds to 9.5 and 10 binary orbits, respectively.  Contours depict approximate density-weighted average temperatures along the line of sight for the H$_2$O (white), NH$_3$ (cyan), and CO$_2$ (black) snow lines at 130 K, 80 K, and 50 K, respectively. The green area represents temperatures below 30 K, visible only in the lower-left panel for these configurations, which traces the CO and N$_2$ snowlines.}
    % ; ./splitpart dump_00100 dumphires_00100 to better show results}}
    \label{fig:Snowlines}
\end{center}
\end{figure*}     

\subsubsection{Evolution of the snow line position}\label{sec:snowline}

\rev{We defined the snow line as the isothermal contour in our density-weighted average temperature maps that matches the freeze-out temperature of a specific molecular species. The complete derivation of the freeze-out temperatures for the molecules considered is provided in Appendix~\ref{app:fo}.} Figure \ref{fig:Snowlines} illustrates the changes in the predicted snow lines for H$_2$O (120 K), NH$_3$ (80 K), CO$_2$ (50 K), CO (30 K), and N$_2$ (30 K) at maximum and minimum stellar separations (periastron and apastron, respectively) for \rev{the Bin-q1} case. We selected H$_2$O, CO, and N$_2$ due to their crucial roles as major volatile carriers in the chemical reactions involving oxygen, carbon, and nitrogen within PPDs \citep{Oberg&Bergin2021}; they regulate the carbon-to-oxygen (C/O) and nitrogen-to-oxygen (N/O) ratios in both gas and solid phases throughout the disk. Additionally, we include CO$_2$ and NH$_3$, which are components of the ice and also influence these ratios \citep{Boogert+2015,McClure+2023}. 
\rev{While we currently employ a single-value approximation for binding energies, we acknowledge that recent theoretical and experimental works \citep[e.g.][]{Minissale+2022, Ferrero+2020} point towards a distribution that can substantially affect snowline location and gas-phase molecular abundance \citep{Tinacci+2023, Boitard-Crepeau+2025}, indicating the need to consider binding energy distributions in future models.}
% It is essential to acknowledge the complexity of determining snow line locations \citep[see][]{Tinacci+2023}; therefore, our proposed snow line positions should be viewed as approximations.

% \nc{Is this expected to happen in more realistic disks in general? I suppose so, you should mention what this could change and for which spectral types, disks radial extension, or sizes this effect could be relevant. OK that this is not expected to happen in this specific case, phantom-mcfost is able to capture this, right? Would it make sense to run a couple of sims for considering a larger value of the binary semi-major axis as a medium separation binary?}

The positions of the snow lines are influenced by both the binary system's orbital phase and the inclination of the disk in relation to the binary's orbital plane. Although the secondary star initially does not possess its own circumstellar disk, tidal forces truncate the primary's disk at approximately $\sim 6$ au \citep[see][]{Pichardo+2005,Miranda+2015}, redistributing material to create a less massive disk around the secondary star. This leads to a more extended H$_2$O and NH$_3$ snow lines surrounding the secondary star, given that both stars share the same stellar characteristics. \rev{The binary gravitational pull is expected to render both disks eccentric, as observed in our simulations. Although the duration of the simulations in our study is too short to make definitive statements about the long-term evolution of the disk structure, it is important to note that if the disk becomes eccentric, the snow lines will also naturally assume an eccentric shape. For instance, the H$_2$O snow line in the coplanar configuration exhibits eccentricities of 0.06 at apocenter and 0.07 at pericenter. In contrast, the inclined configuration shows an eccentricity of 0.06 at apocenter but a reduced value of 0.02 at pericenter. 
} However, the relatively high temperature (120 K) of the H$_2$O snow line makes its position less sensitive to changes in binary-disk inclination within the parameters considered in these simulations, keeping its location close to the stars. Conversely, a larger luminosity ratio favouring the secondary star or a less massive circumprimary disk could significantly affect the position of the H$_2$O snow line. Such an inclination-shift dependence is more readily visible for the CO$_2$ snowline. The CO$_2$ snow line in these simulations closely follows regions of high optical depth, indicating areas with sufficiently dense gas to maintain low temperatures. This snow line is mainly observed around the circumprimary disk.

\begin{figure}
\centering
\begin{center}
    \includegraphics[width=0.5\textwidth]{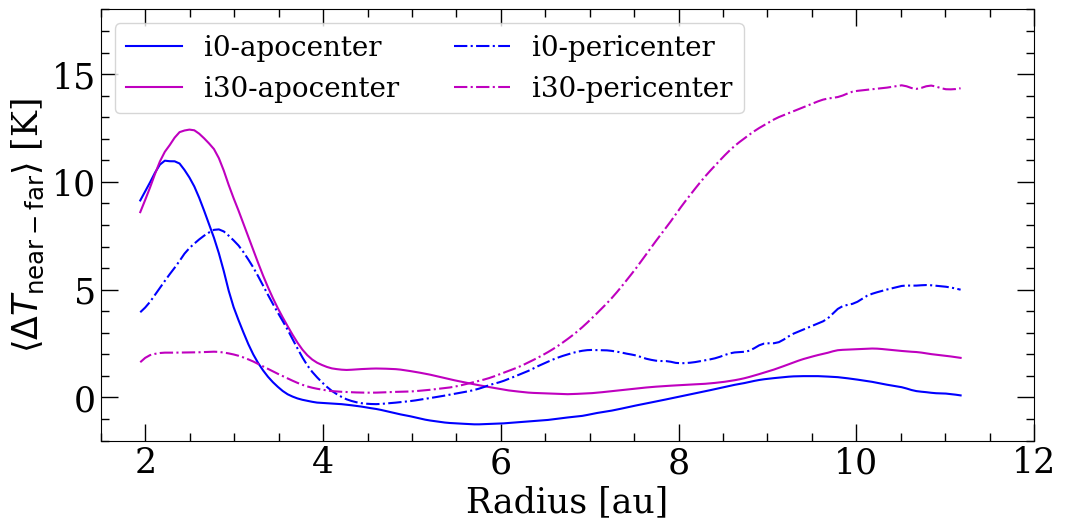} 
    \caption{\rev{Density-weighted average temperature difference between the near and far sides of the disk in the Bin-q1 simulation shown as a function of radius for both the coplanar (blue) and inclined (magenta) configurations. The solid lines indicate the apocenter position of the binary orbit, while the dashed-dotted lines correspond to the pericenter.}}
    \label{fig:Tb}
\end{center}
\end{figure}     

Coplanar configurations maximize optical depth along the disk-plane for the secondary star's radiation, resulting in less efficient heating of the circumprimary disk compared to inclined configurations. Consequently, the coplanar simulations display the coldest disk temperatures ($<30$ K). The companion star's gravitational influence at the pericentre triggers prominent gas spiral arms. The CO$_2$ snow line then traces these spirals, encompassing regions with the lowest temperatures below 30 K on the opposite side of the companion's location, indicating frozen CO and N$_2$. Although the spiral arm nearest the companion star is denser, it experiences more significant heating from the companion, slightly pushing the snow line back. In contrast, the CO$_2$ snow line remains nearly symmetric and mildly eccentric around the primary star at periastron, within an inner radius lower than 7 au. However, in the inclined case, \rev{several azimuthal directions facing the companion reach temperatures exceeding 50 K}, causing the CO$_2$ snow line to lose its radial symmetry across all radii.
% Additionally, the NH$_3$ snow line occurs at a distance of nearly 7 au.
Figure \ref{fig:Tb} \rev{shows this asymmetric profile; it} shows a temperature difference of almost 15 K between the disk's two sides at pericenter in the inclined configuration, which largely explains the shape of the CO$_2$ snow line.
%
% The loss of symmetry and the snowline's dependence on the binary phase significantly affect the dust grain and disk chemistry in PPDs. In the coplanar configuration, the chemistry is primarily governed by the coldest regions, where lower temperatures can cause volatile compounds to accumulate on dust grains. Conversely, additional heating modifies the chemistry in the inclined configuration by potentially desorbing these compounds and impacting the disk's overall temperature profile. These changes can influence the formation and distribution of complex molecules within the disk. The detailed implications of these phenomena on the evolution of the disk will be thoroughly examined in Section \ref{sec:diskussion}.

\subsection{Effects of dust settling on the mid-plane}
\label{sec:dust_rad_comp}

\begin{figure*}
\centering
\begin{center}
    \includegraphics[width=1.0\textwidth]{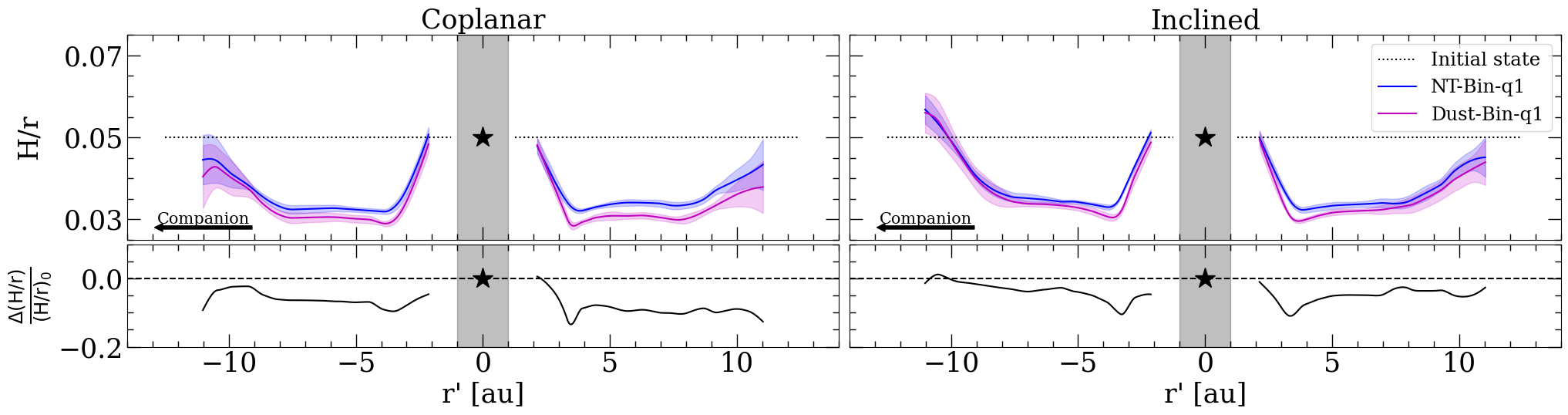} 
    \caption{Disk aspect ratio profiles \rev{and relative variations} from \rev{NT-Bin-q1} (blue) and \rev{Dust-Bin-q1} (magenta) simulations for the Bin-q1 case. The bottom panels quantify the relative difference in aspect ratio, $\Delta (H/r)$, between these two simulations, normalized to the aspect ratio from the dust-free simulation, $(H/r)_0$.}
    \label{fig:dust_rad_comp}
\end{center}
\end{figure*}

In contrast to the dust-free simulation \rev{(NT-Bin-q1)}, the one-fluid dust simulation \rev{(Dust-Bin-q1)} does not assume a constant dust-to-gas ratio for all the considered evolutionary time; rather, it explicitly models dust evolution, with dust settling towards the mid-plane as expected \citep{Price&Laibe2015,Hutchison+2018}. The accumulation of dust at the disk mid-plane increases the optical depth and reduces the temperature, consequently decreasing the aspect ratio, as illustrated in Figure \ref{fig:dust_rad_comp}. 
% \cp{I would be careful on any interpretation here, as you only have 3 grain sizes, this is unlikely to be enough to get an accurate dust distribution for mcfost}

The change in aspect ratio is a function of binary inclination. In the coplanar \rev{Dust-Bin-q1 case}, the reduction is notably more pronounced on the side opposite the companion star, displaying a decrease of approximately 15\% compared to the \rev{NT-Bin-q1 simulation}. 
% \cp{I am not convinced this is correct as you do not model the small grains which are responsible for the bulk of the opacity in the optical near-IR} 
The companion's heating causes a comparatively higher aspect ratio on the side facing it. Thus, the difference in aspect ratio progressively increases between these two sides from nearer to farther side to the companion (in regions with $\left|r'\right|>3$ au) because the far side is not affected by the companion's radiation (see Figure \ref{fig:hr_sigma_prof}).

In the inclined \rev{Dust-Bin-q1 case}, aspect ratio variations stay below 5\%. The most significant decreases in aspect ratio occur at the location of the dust inner rim ($\sim3$ au), a feature also observed in the coplanar \rev{Dust-Bin-q1 simulation}. This aligns with a dust-dense region where the temperature is expected to be lower. The disk side facing the companion presents the same aspect ratio between the two dust prescriptions. This indicates that the settled mid-plane dust primarily contributes to extinguishing all radiation penetrating along the disk mid-plane direction, consistent with the optical depth behavior. 

\subsection{Unequal mass binary and outburst event}
\label{sec:Fuori}

\begin{figure*}
\centering
\begin{center}
    \includegraphics[width=1.0\textwidth]{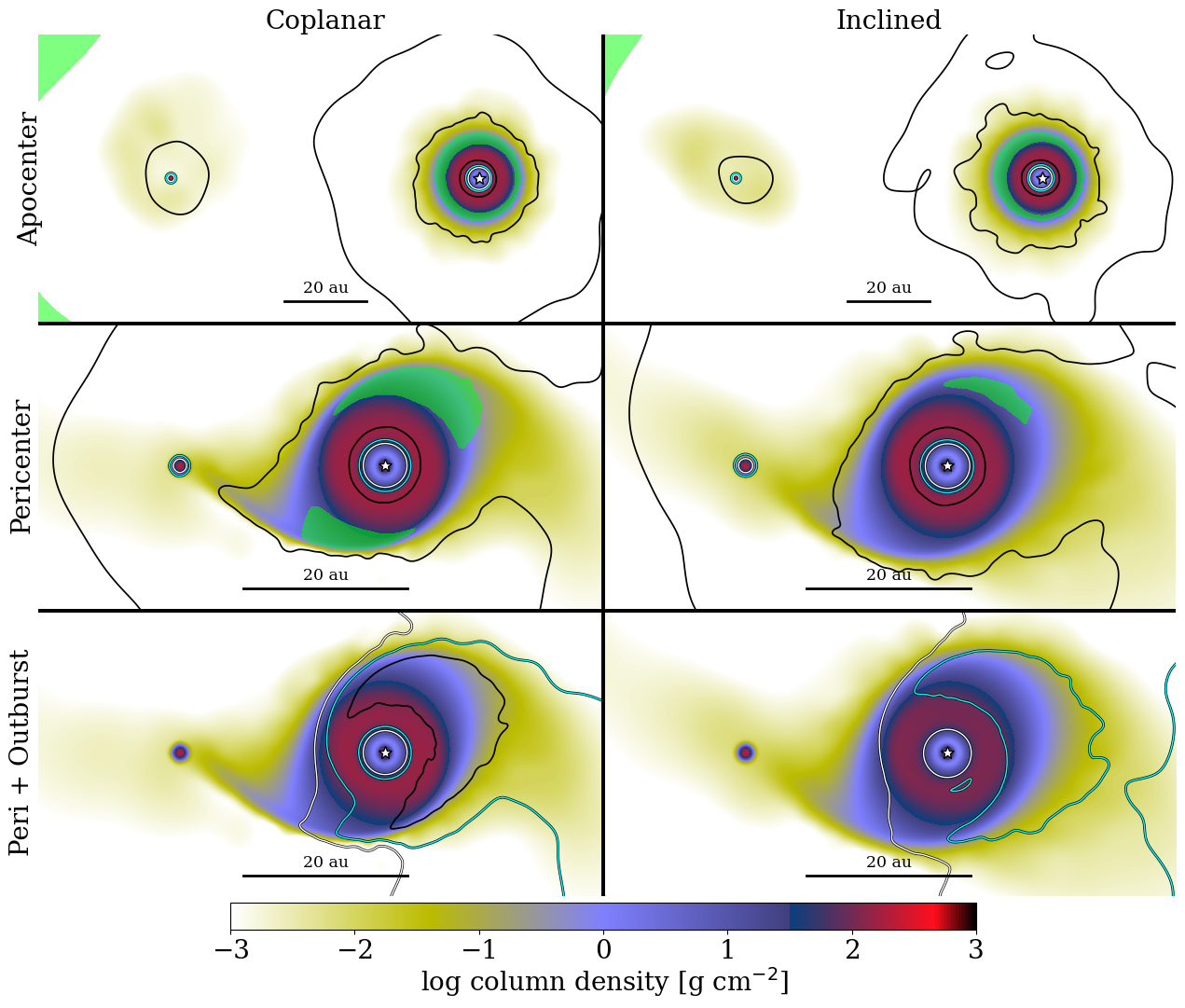} 
    \caption{Column density maps of circumstellar disks for Bin-q05 and FU-Bin-q05, labelled similarly to Figure \ref{fig:Snowlines}. \rev{The green area represents temperatures below 30 K}. The upper and middle rows illustrate the disk morphology of Bin-q05 at apocentre and pericentre, respectively. The bottom row depicts the same evolutionary time shown in the middle row, but for the outbursting companion star: FU-Bin-q05. The secondary star is not depicted for easy visualization of its circumstellar disk.}
    \label{fig:Snowlines*}
\end{center}
\end{figure*}

Figure \ref{fig:Snowlines*} presents a comparison similar to that of Figure \ref{fig:Snowlines}, but specifically for Bin-q05 and FU-Bin-q05. Given that the binary mass ratio is 0.5, the truncation radius for the CSD is larger than that of Bin-q1, allowing more material from the CSD to remain bound to the primary star. This increases the disk's optical thickness, which, together with a less luminous companion, results in an overall colder disk. This phenomenon is evident in both the coplanar and inclined configurations, where the outermost disk regions exhibit temperatures below 30 K, \rev{which we have defined as green areas}, during the apocentre binary phase. In contrast to the inclined Bin-q1 simulation at apocentre, the CSD in inclined Bin-q05 shows that the isothermal lines for 30 K and 50 K are more extended toward the location of the companion star and narrower on the opposite disk side. This thermal structure can be attributed to the low luminosity of the companion. Despite the presence of a moderate level of disk eccentricity, similar to that in Bin-q1, the companion's luminosity, approximately 30 times less than that of the primary star, prevents it from effectively heating the outer regions of the CSD.

During the pericentre binary phase in the coplanar Bin-q05 simulation, the green areas are not only more prominent but also extend into another disk sector. In the inclined setup, the green region appears in the same location as in the coplanar Bin-q1 case but is slightly larger. Additionally, the CO$_2$ snow line remains continuous azimuthally, unlike the discontinuities observed in inclined Bin-q1. These results underscore the importance of the disk’s optical properties, in conjunction with the stellar properties, in determining the disk thermal structure. 

The bottom row of Figure \ref{fig:Snowlines*} depicts the effects of the outbursting companion represented with the FU-Bin-q05 simulation. The abrupt increase in the companion's luminosity significantly changes the temperature profile of the entire CSD of the primary star. 

In the coplanar \rev{FU-Bin-q05} simulation, the H$_2$O and NH$_3$ snow lines shift from an initial position near the secondary star (pre-outburst) to the outer regions of the primary CSD. The CO$_2$ snow line experiences a change similar to that seen in inclined Bin-q1 at pericentre; it loses its azimuthal symmetry, with the areas facing the companion heating to temperatures exceeding 50 K. Notably, the NH$_3$ and CO$_2$ snow lines display comparable behaviour in both coplanar FU-Bin-q05 and inclined Bin-q1 cases. This indicates that an inclined binary can cause similar heating effects in a disk as a coplanar disk undergoing external heating due to a stellar outburst for certain isothermal lines. On the other hand, in the inclined FU-Bin-q05 simulation, the disk's temperature exceeds 50 K, indicating the complete sublimation of CO$_2$. Additionally, the shape of the NH$_3$ snow line closely resembles that of the CO$_2$ snow line in the coplanar FU-Bin-q05 simulation.

\begin{figure*}
\centering
\begin{center}
    \includegraphics[width=1.0\textwidth]{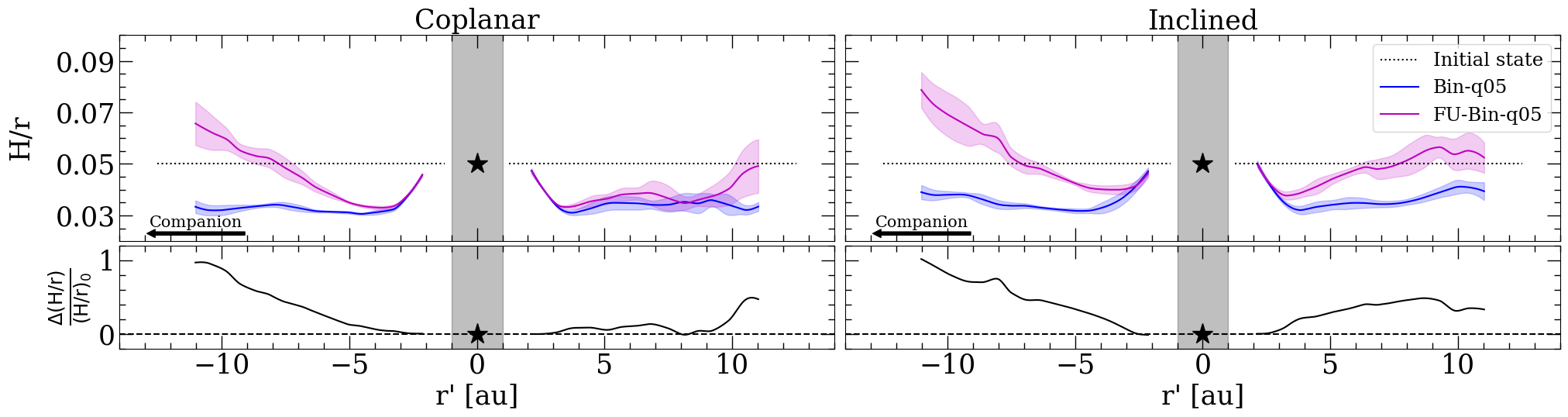} \\
    \caption{Figure compares the disk aspect ratio profile for Bin-q05 (blue) and FU-Bin-q05 (magenta). The bottom panels quantify the relative differences between these two quantities across the two simulations, normalized to the values obtained from Bin-q05.}
    \label{fig:rad_comp*}
\end{center}
\end{figure*}

\rev{Figure~\ref{fig:rad_comp*} depicts the dynamical aftermath of the outburst over the disk.} The curves represent the median value for 43 years following the outburst, instead of the entire orbit as shown in Figure~\ref{fig:hr_sigma_prof}. In both coplanar and inclined \rev{FU-Bin-q05 simulations}, the disk's side facing the companion experiences a 100\% increase in the aspect ratio value compared to the pre-outburst value. The opposite side of the disk experiences variation in regions with $r'<10$ au of no more than 20\% \rev{for the coplanar setup}. In the inclined case, however, the aspect ratio increases up to 50\%. 
% \pp{Does the density panel contribute anything? See Figure \ref{fig:SD_out}}. \nc{This last subsection is excellent! Not sure the density adds much to the story. If needed, we can always show in the complementary material.}

\section{Discussion}
\label{sec:diskussion}

\rev{This new {\texttt Phantom}-{\texttt Mcfost} implementation allowed us to explore the radiative feedback and circumstellar disk morphology induced by a bound stellar companion.} Our results highlight the importance of accounting for radiative processes, as they affect two main areas: dynamical behavior and astrochemical evolution. These aspects play a crucial role in imprinting observable features within binary stellar systems.

% \subsection{Dynamical implications}
% \label{sec:Dyn_implications}

% The aspect ratio can vary freely based on the local thermal properties to achieve hydrostatic equilibrium in our simulations. As discussed in Section \ref{sec:iso-rad_comp}, a stellar companion can cause the outer regions of a CSD to puff up vertically. This effect may result in the primary star casting shadows in a circumbinary disk, if present, with shadows displaying different shapes due to the asymmetric light absorption of the CSD. \cp{This is a bit speculative}  Shadows are not uncommon in PPDs; \citet{Bohn+2022} extensively explore this phenomenon across various systems. Notably, we highlight the binary system HD~142527 \citep{Marino+2015}, the triple system GW~Orionis \citep{Kraus+2020}, and the quadruple stellar system GG~Tau \citep{Keppler2020} as prominent examples of such shadowing effects in multiple stellar systems.

The aspect ratio affects grain growth and planetary migration \citep{Brauer+2008}. Despite the rapid accretion of gas and dust from the CSD into the primary star \citep{Monin+2007}, the presence of a CBD can act as a reservoir of material to replenish the CSD \citep{Nelson&Marzani2016, Marzari&DAngelo2025}. In these long-lived CSDs, dust grains can evolve not only under the companion's gravitational influence but also due to its radiative effects. When planets finally form, their evolution remains connected to the $H/r$ value through the disk's torque \citep[see][]{Paardekooper+2023}. Then, an asymmetric aspect ratio along a fixed radial distance will lead to differential migration rates along the planetary orbit.

% \nc{We should probably establish a link with the constraints obtained by the edge-on disks programme by F et al. If my memory is correct, there is one system for which binarity is suspected (unless it is a projection effect). Are there any variations of H/r observed there? If not, your results could help to provide more support to one of the two scenarios. Assuming there is no observational evidence of this effect at the moment, you could roughly estimate the expected vertical variations as a function of the wavelength for a typical disk or one of your examples. This would allow you to estimate the resolution required to actually observe, which could in turn help for future ALMA proposal. Last, long term monitoring (let's say half the binary period) could also help to spot vertical and azimuthal variability. I would briefly mention this if relevant.}
 
% \subsection{Astrochemical implications}
% \label{sec:Chem_implications}

Shifting snow lines influence the availability of key species in both the gas and solid phases, directly impacting the C/O and N/O ratios in various disk regions. These changes in chemical abundances can affect the formation and composition of planetesimals and, ultimately, the atmospheres of emerging planets \citep{Oberg+2011,Oberg+2016, Booth+2017, Madhusudhan2019}. For instance, the increased heating from the secondary star in inclined systems may lead to the sublimation of volatile compounds from dust grains, enriching the gas phase and altering the overall chemical environment. Furthermore, the presence of spiral arms induced by the companion star can create localized areas of high density and temperature gradients. 
\rev{The sublimation of molecules in the gas phase alters the distribution of pre-existing species and promotes the formation of interstellar complex organic molecules \citep[COMs;][]{Herbst&vanDishoeck2009,Ceccarelli+2023}.}

\begin{figure*}
\centering
\begin{center}
    \includegraphics[width=1.0\textwidth]{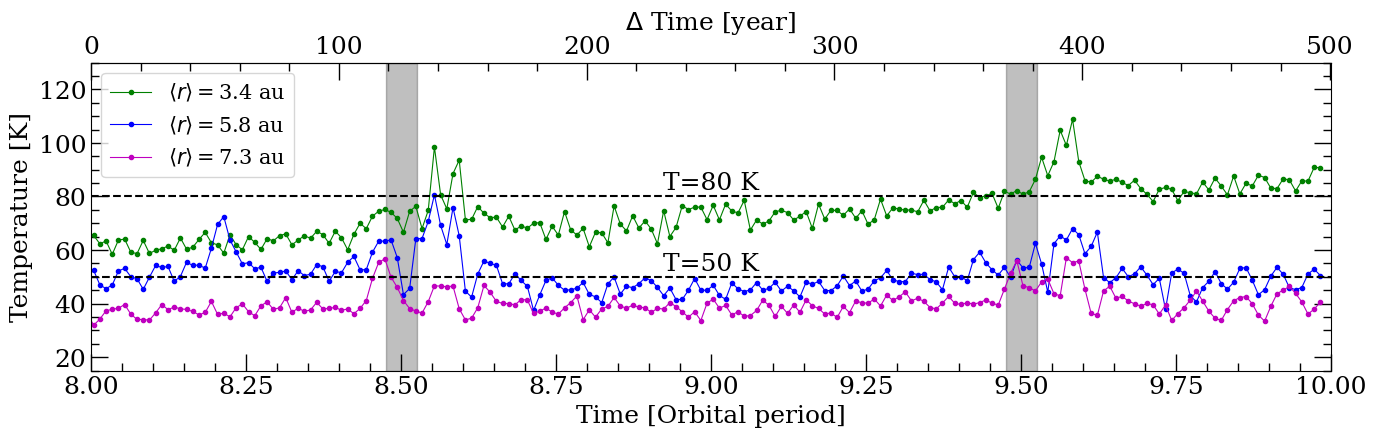} \\
    \caption{Figure illustrates the temperature evolution, \rev{for the inclined Bin-q1 simulation}, along the two final binary orbits for three SPH particles, which represent different mean radial distances: 3.4 au (green), 5.8 au (blue), and 7.3 au (magenta). The vertical gray areas highlight the pericenter phase of the binary system, while the horizontal dashed lines indicate two specific temperature values.}
    \label{fig:tracking}
\end{center}
\end{figure*}

The additional heating from the secondary star and the periodical formation of spiral arms affect the CSD evolution. This translates into thermal cycles where particles undergo phases of high temperatures (at pericentre) and maintain a relatively constant low temperature during the rest of the binary orbit. Figure \ref{fig:tracking} illustrates the temperature modulation for three \rev{representative} SPH particles located at different radii in the inclined \rev{Bin-q1 simulation}. Note that the peak temperature is shifted from the binary's pericentre due to the particle's differential rotation, as explained in Section \ref{sec:ph+mc}. Typically, PPDs around late Class I or Class II single stars exhibit a lack of emission in COMs due to their cooler disks compared to their Class~0 \rev{or early Class~I} counterparts \citep{Villarmois+2019, Hoff+2020}. This limitation can be alleviated by considering the heating caused by a companion star, particularly relevant in inclined configurations, as discussed in Section \ref{sec:snowline}. Therefore, binary systems present a valuable opportunity to explore volatile molecules in evolved systems. Specifically, binary systems may act as a significant source of abundant COM emissions in \rev{late} Class~I or Class~II stars, contrasting with single stellar systems, where COMs are typically frozen onto ice grains due to lower temperatures.

Although fluctuations in particle temperature correlate with the companion's position, this behavior is primarily driven by the LTE assumption in the cell temperature calculation. Sublimation of species from dust grains occurs nearly instantaneously when sufficient energy is supplied. However, freeze-out or adsorption, the process of converting volatile species back to ice, exhibits a delay following luminosity changes, as reported by \citet{Lee+2025}. Consequently, particles can remain in the gaseous phase even after the luminosity source has passed. Appendix \ref{app:fo} outlines the relationship between the freeze-out timescale and the disk's local conditions. In our simulations, the number density of hydrogen nuclei ($n_{\rm H}$) in the CSD mid-plane is approximately $10^{12}\ \rm cm^{-3}$, resulting in a freeze-out timescale of slightly more than two days for CO$_2$ according to Equation \ref{eq:time_fo}. Even applying a decrease of three orders of magnitude for the value of $n_{\rm H}$, the freeze-out timescale remains below $10$ years. Therefore, this astrochemical process could be readily traced on timescales from days to years. Similar to the B335 system, binary stars may provide another potential natural laboratory for astrochemistry \citep{Lee+2025}. 

\rev{To observe this process, binary systems should} possess partially resolved CSDs that enable the detection of azimuthal intensity variations. Although larger CSDs are easier to characterize, they often signify distant companions with long orbital periods. We emphasize the complex interplay of stellar parameters, as disk size is dependent on the binary mass ratio \citep{Pichardo+2005}, and stellar mass affects the star's temperature. The key parameters of the companion include its luminosity, inclination, and proximity to the disk. Our results indicate that temperature variations are observable in disks approximately 10 au in radius around Sun-like stars, especially in inclined configurations. Additionally, binary-disk inclination tends to lead to radial disk widening, with the most pronounced effects observed in retrograde orbits \citep{Miranda+2015}. Following these and our findings, binaries with similar stellar masses, semi-major axes less than $\sim 50$ au, and eccentricities below 0.5, featuring bright and inclined companions, would be favourable targets for studying volatile molecules across binary phases. 

\rev{Finally, the age of the binary system also affects the thermal structure of the disk. In the Class~0 or early Class~I phase, the surrounding envelope both heats the system and shields the disks from direct irradiation by the companion, while abundant material leads to frequent outbursts. As the envelope dissipates in later stages, stellar luminosity and disk mass decline, and the disks become directly exposed to irradiation, resulting in thermal patterns that differ from those in the embedded phase.} The outburst event driven by episodic accretion amplifies the thermal influence of the companion star on the CSD, with FU-Orionis and EXor types being the most characteristic \citep{Hartmann&Kenyon1996,Audard+2014}. Binary stars increase the likelihood of these outbursts at the pericentre, particularly when they exhibit eccentricity \citep{Bonnell&Bastien1992}; the disks surrounding both stars \rev{develop} spiral waves that \rev{trigger accretion onto} the stars \citep{Vorobyov&Basu2005}. Although both stars can undergo accretion, an outburst from an inclined secondary relative to the CSD will cause the most significant changes in the chemical structure of the disk. Additionally, \citet{Ros&Johansen2024} investigated planetesimal formation and growth under the influence of stellar outbursts, demonstrating efficient grain growth up to centimetre sizes. Consequently, the periodic behaviour of the binary system in triggering outbursts at pericentre could enhance the formation of planetesimals.

\section{Conclusion}
\label{sec:conclusion}

We conducted hydrodynamic simulations of a CSD under the gravitational and radiative influence of a stellar binary system. We explored various disk-binary orientations, binary mass ratios, and dust prescriptions, and we simulated an FU-Orionis-type outburst in the secondary star. Our results reveal a strong dependence among the properties of the PPD, stellar characteristics, disk-binary inclination, and binary orbital phase in heating different regions of the disk. These are our main findings:
\begin{enumerate}[(i)]

    \item The addition of a radiative transfer solver in a 3D hydrodynamical simulation enhances the temperature profile of the PPD by considering the disk's optical, thermal, and chemical properties alongside the stellar properties. Consequently, the efficiency of dust settling becomes an important factor in determining the extinction levels along the disk mid-plane.

    \item The inclusion of a secondary star introduces significant asymmetries in disk temperature profiles, affecting both the vertical structure and the locations of the snow lines. The disk-binary inclination and binary phase strongly modulate the disk's temperature profile, with inclined configurations efficiently heating the disk mid-plane. The outburst amplifies the radiative effects observed in the quiescent stellar state compared to the CSD.
    
    \item The asymmetrical disk's temperature profile leads to differential snow line profiles for various species. Molecules can sublimate and freeze out along a single orbit, which can be observed due to the freeze-out timescales of less than 10 years under typical PPD conditions. These changes in the snow lines can significantly influence the chemical environment and material availability for planet formation, differing from those in single-star systems.
\end{enumerate}

Our work represents a significant advancement in addressing various astrophysical problems by incorporating radiative transfer calculations alongside 3D hydrodynamical modeling for PPDs in stellar binaries. This implementation provides predictions of observable signposts that can be tracked over time in a CSD, enabling a deeper understanding of the interactions between physical and chemical processes in PPDs. Future monitoring of disks in binary systems, particularly those experiencing episodic accretion events, will be crucial. Such observations will help constrain not only the dynamical aspects of these complex systems but also yield insights into their chemical composition and structure. Understanding these factors will ultimately enhance our knowledge of planet formation processes, the environmental conditions that influence the development of planetary systems across binary stellar systems, and how they differ from single systems.

\begin{acknowledgements} 
This project has received funding from the European Research Council (ERC) under the European Union Horizon Europe programme (grant agreement No. 101042275, project Stellar-MADE).
\end{acknowledgements}

\section*{Data availability}
The data underlying this article will be shared at a reasonable request by the corresponding author. 

% We used the following public software:\\ 
% \newline
% \begin{tabular}{ll}
%     \texttt{Phantom } & \hyperlink{https://github.com/danieljprice/phantom}{https://github.com/danieljprice/phantom} \\
%      &  \citet{PricePH+2018b} \\
%      \texttt{Mcfost} & \hyperlink{https://github.com/cpinte/mcfost}{https://github.com/cpinte/mcfost} \\
%      &  \citet{Pinte+2006,Pinte+2009} \\
%      \texttt{Splash} & \hyperlink{https://github.com/danieljprice/splash}{https://github.com/danieljprice/splash} \\
%      &  \citet{Price2007}
% \end{tabular}

\bibliographystyle{aa}
\bibliography{paper}

\begin{thebibliography}{88}
\expandafter\ifx\csname natexlab\endcsname\relax\def\natexlab#1{#1}\fi

\bibitem[{{Armitage}(2018)}]{Armitage2018}
{Armitage}, P.~J. 2018, in Handbook of Exoplanets, ed. H.~J. {Deeg} \& J.~A. {Belmonte}, 135

\bibitem[{{Artur de la Villarmois} {et~al.}(2019){Artur de la Villarmois}, {J{\o}rgensen}, {Kristensen}, {Bergin}, {Harsono}, {Sakai}, {van Dishoeck}, \& {Yamamoto}}]{Villarmois+2019}
{Artur de la Villarmois}, E., {J{\o}rgensen}, J.~K., {Kristensen}, L.~E., {et~al.} 2019, \aap, 626, A71

\bibitem[{{Artymowicz} \& {Lubow}(1994)}]{Artymowicz&Lubow1994}
{Artymowicz}, P. \& {Lubow}, S.~H. 1994, \apj, 421, 651

\bibitem[{{Audard} {et~al.}(2014){Audard}, {{\'A}brah{\'a}m}, {Dunham}, {Green}, {Grosso}, {Hamaguchi}, {Kastner}, {K{\'o}sp{\'a}l}, {Lodato}, {Romanova}, {Skinner}, {Vorobyov}, \& {Zhu}}]{Audard+2014}
{Audard}, M., {{\'A}brah{\'a}m}, P., {Dunham}, M.~M., {et~al.} 2014, in Protostars and Planets VI, ed. H.~{Beuther}, R.~S. {Klessen}, C.~P. {Dullemond}, \& T.~{Henning}, 387--410

\bibitem[{{Ballabio} {et~al.}(2018){Ballabio}, {Dipierro}, {Veronesi}, {Lodato}, {Hutchison}, {Laibe}, \& {Price}}]{Ballabio+2018}
{Ballabio}, G., {Dipierro}, G., {Veronesi}, B., {et~al.} 2018, \mnras, 477, 2766

\bibitem[{{Baraffe} {et~al.}(2015){Baraffe}, {Homeier}, {Allard}, \& {Chabrier}}]{Baraffe+2015}
{Baraffe}, I., {Homeier}, D., {Allard}, F., \& {Chabrier}, G. 2015, \aap, 577, A42

\bibitem[{{Bate} {et~al.}(1995){Bate}, {Bonnell}, \& {Price}}]{Bate+1995}
{Bate}, M.~R., {Bonnell}, I.~A., \& {Price}, N.~M. 1995, \mnras, 277, 362

\bibitem[{{Boitard-Cr{\'e}peau} {et~al.}(2025){Boitard-Cr{\'e}peau}, {Ceccarelli}, {Beck}, {Vacher}, \& {Ugliengo}}]{Boitard-Crepeau+2025}
{Boitard-Cr{\'e}peau}, L., {Ceccarelli}, C., {Beck}, P., {Vacher}, L., \& {Ugliengo}, P. 2025, \apjl, 987, L25

\bibitem[{{Bonnell} \& {Bastien}(1992)}]{Bonnell&Bastien1992}
{Bonnell}, I. \& {Bastien}, P. 1992, \apjl, 401, L31

\bibitem[{{Boogert} {et~al.}(2015){Boogert}, {Gerakines}, \& {Whittet}}]{Boogert+2015}
{Boogert}, A.~C.~A., {Gerakines}, P.~A., \& {Whittet}, D. C.~B. 2015, \araa, 53, 541

\bibitem[{{Booth} {et~al.}(2017){Booth}, {Clarke}, {Madhusudhan}, \& {Ilee}}]{Booth+2017}
{Booth}, R.~A., {Clarke}, C.~J., {Madhusudhan}, N., \& {Ilee}, J.~D. 2017, \mnras, 469, 3994

\bibitem[{{Borchert} {et~al.}(2022{\natexlab{a}}){Borchert}, {Price}, {Pinte}, \& {Cuello}}]{Borchert+2022a}
{Borchert}, E. M.~A., {Price}, D.~J., {Pinte}, C., \& {Cuello}, N. 2022{\natexlab{a}}, \mnras, 510, L37

\bibitem[{{Borchert} {et~al.}(2022{\natexlab{b}}){Borchert}, {Price}, {Pinte}, \& {Cuello}}]{Borchert+2022b}
{Borchert}, E. M.~A., {Price}, D.~J., {Pinte}, C., \& {Cuello}, N. 2022{\natexlab{b}}, \mnras, 517, 4436

\bibitem[{{Boyajian} {et~al.}(2012){Boyajian}, {von Braun}, {van Belle}, {McAlister}, {ten Brummelaar}, {Kane}, {Muirhead}, {Jones}, {White}, {Schaefer}, {Ciardi}, {Henry}, {L{\'o}pez-Morales}, {Ridgway}, {Gies}, {Jao}, {Rojas-Ayala}, {Parks}, {Sturmann}, {Sturmann}, {Turner}, {Farrington}, {Goldfinger}, \& {Berger}}]{Boyajian+2012}
{Boyajian}, T.~S., {von Braun}, K., {van Belle}, G., {et~al.} 2012, \apj, 757, 112

\bibitem[{{Brauer} {et~al.}(2008){Brauer}, {Dullemond}, \& {Henning}}]{Brauer+2008}
{Brauer}, F., {Dullemond}, C.~P., \& {Henning}, T. 2008, \aap, 480, 859

\bibitem[{{Ceccarelli} {et~al.}(2023){Ceccarelli}, {Codella}, {Balucani}, {Bockelee-Morvan}, {Herbst}, {Vastel}, {Caselli}, {Favre}, {Lefloch}, {Oberg}, \& {Yamamoto}}]{Ceccarelli+2023}
{Ceccarelli}, C., {Codella}, C., {Balucani}, N., {et~al.} 2023, in Astronomical Society of the Pacific Conference Series, Vol. 534, Protostars and Planets VII, ed. S.~{Inutsuka}, Y.~{Aikawa}, T.~{Muto}, K.~{Tomida}, \& M.~{Tamura}, 379

\bibitem[{{Chiang} \& {Goldreich}(1997)}]{Chiang&Goldraich1997}
{Chiang}, E.~I. \& {Goldreich}, P. 1997, \apj, 490, 368

\bibitem[{{Cieza} {et~al.}(2016){Cieza}, {Casassus}, {Tobin}, {Bos}, {Williams}, {Perez}, {Zhu}, {Caceres}, {Canovas}, {Dunham}, {Hales}, {Prieto}, {Principe}, {Schreiber}, {Ruiz-Rodriguez}, \& {Zurlo}}]{Cieza+2016}
{Cieza}, L.~A., {Casassus}, S., {Tobin}, J., {et~al.} 2016, \nat, 535, 258

\bibitem[{{Cridland} {et~al.}(2020){Cridland}, {van Dishoeck}, {Alessi}, \& {Pudritz}}]{Cridland+2020}
{Cridland}, A.~J., {van Dishoeck}, E.~F., {Alessi}, M., \& {Pudritz}, R.~E. 2020, \aap, 642, A229

\bibitem[{Cuello {et~al.}(2025)Cuello, Alaguero, \& Poblete}]{Cuello+2025}
Cuello, N., Alaguero, A., \& Poblete, P.~P. 2025, Symmetry, 17

\bibitem[{{Duch{\^e}ne} \& {Kraus}(2013)}]{Duchene&Kraus2013}
{Duch{\^e}ne}, G. \& {Kraus}, A. 2013, \araa, 51, 269

\bibitem[{{Dunhill} {et~al.}(2015){Dunhill}, {Cuadra}, \& {Dougados}}]{Dunhill+2015}
{Dunhill}, A.~C., {Cuadra}, J., \& {Dougados}, C. 2015, \mnras, 448, 3545

\bibitem[{{Facchini} {et~al.}(2013){Facchini}, {Lodato}, \& {Price}}]{Facchini+2013}
{Facchini}, S., {Lodato}, G., \& {Price}, D.~J. 2013, \mnras, 433, 2142

\bibitem[{{Ferrero} {et~al.}(2020){Ferrero}, {Zamirri}, {Ceccarelli}, {Witzel}, {Rimola}, \& {Ugliengo}}]{Ferrero+2020}
{Ferrero}, S., {Zamirri}, L., {Ceccarelli}, C., {et~al.} 2020, \apj, 904, 11

\bibitem[{{Fischer} {et~al.}(2023){Fischer}, {Hillenbrand}, {Herczeg}, {Johnstone}, {Kospal}, \& {Dunham}}]{Fischer+2023}
{Fischer}, W.~J., {Hillenbrand}, L.~A., {Herczeg}, G.~J., {et~al.} 2023, in Astronomical Society of the Pacific Conference Series, Vol. 534, Protostars and Planets VII, ed. S.~{Inutsuka}, Y.~{Aikawa}, T.~{Muto}, K.~{Tomida}, \& M.~{Tamura}, 355

\bibitem[{{Franchini} {et~al.}(2019){Franchini}, {Martin}, \& {Lubow}}]{Franchini+2019}
{Franchini}, A., {Martin}, R.~G., \& {Lubow}, S.~H. 2019, \mnras, 485, 315

\bibitem[{{Fu} {et~al.}(2015){Fu}, {Lubow}, \& {Martin}}]{Fu+2015}
{Fu}, W., {Lubow}, S.~H., \& {Martin}, R.~G. 2015, \apj, 807, 75

\bibitem[{{Gundlach} \& {Blum}(2015)}]{Gundlach&Blum+2015}
{Gundlach}, B. \& {Blum}, J. 2015, \apj, 798, 34

\bibitem[{{Hartmann} \& {Kenyon}(1996)}]{Hartmann&Kenyon1996}
{Hartmann}, L. \& {Kenyon}, S.~J. 1996, \araa, 34, 207

\bibitem[{{Herbst} \& {van Dishoeck}(2009)}]{Herbst&vanDishoeck2009}
{Herbst}, E. \& {van Dishoeck}, E.~F. 2009, \araa, 47, 427

\bibitem[{{Hirsh} {et~al.}(2020){Hirsh}, {Price}, {Gonzalez}, {Ubeira-Gabellini}, \& {Ragusa}}]{Hirsh+2020}
{Hirsh}, K., {Price}, D.~J., {Gonzalez}, J.-F., {Ubeira-Gabellini}, M.~G., \& {Ragusa}, E. 2020, \mnras, 498, 2936

\bibitem[{{Hollenbach} {et~al.}(2009){Hollenbach}, {Kaufman}, {Bergin}, \& {Melnick}}]{Hollenbach+2009}
{Hollenbach}, D., {Kaufman}, M.~J., {Bergin}, E.~A., \& {Melnick}, G.~J. 2009, \apj, 690, 1497

\bibitem[{{Hutchison} {et~al.}(2018){Hutchison}, {Price}, \& {Laibe}}]{Hutchison+2018}
{Hutchison}, M., {Price}, D.~J., \& {Laibe}, G. 2018, \mnras, 476, 2186

\bibitem[{{Kenyon} \& {Hartmann}(1987)}]{Kenyon&Hartmann1987}
{Kenyon}, S.~J. \& {Hartmann}, L. 1987, \apj, 323, 714

\bibitem[{{Kley} \& {Nelson}(2008)}]{Kley+2008a}
{Kley}, W. \& {Nelson}, R.~P. 2008, \aap, 486, 617

\bibitem[{{Kley} {et~al.}(2008){Kley}, {Papaloizou}, \& {Ogilvie}}]{Kley+2008b}
{Kley}, W., {Papaloizou}, J.~C.~B., \& {Ogilvie}, G.~I. 2008, \aap, 487, 671

\bibitem[{{Laibe} \& {Price}(2014{\natexlab{a}})}]{Laibe&Price2014b}
{Laibe}, G. \& {Price}, D.~J. 2014{\natexlab{a}}, \mnras, 444, 1940

\bibitem[{{Laibe} \& {Price}(2014{\natexlab{b}})}]{Laibe&Price2014a}
{Laibe}, G. \& {Price}, D.~J. 2014{\natexlab{b}}, \mnras, 440, 2147

\bibitem[{{Larwood} {et~al.}(1996){Larwood}, {Nelson}, {Papaloizou}, \& {Terquem}}]{Larwood1996}
{Larwood}, J.~D., {Nelson}, R.~P., {Papaloizou}, J.~C.~B., \& {Terquem}, C. 1996, \mnras, 282, 597

\bibitem[{{Lee} {et~al.}(2025){Lee}, {Evans}, {Baek}, {Kim}, {Noh}, \& {Yang}}]{Lee+2025}
{Lee}, J.-E., {Evans}, N.~J., {Baek}, G., {et~al.} 2025, \apjl, 978, L3

\bibitem[{{Lee} {et~al.}(2019){Lee}, {Lee}, {Baek}, {Aikawa}, {Cieza}, {Yoon}, {Herczeg}, {Johnstone}, \& {Casassus}}]{Lee+2019}
{Lee}, J.-E., {Lee}, S., {Baek}, G., {et~al.} 2019, Nature Astronomy, 3, 314

\bibitem[{{Lindblad}(1941)}]{Lindblad1941}
{Lindblad}, B. 1941, Stockholms Observatoriums Annaler, 13, 10.1

\bibitem[{{Lodato} \& {Price}(2010)}]{Lodato&Price2010}
{Lodato}, G. \& {Price}, D.~J. 2010, \mnras, 405, 1212

\bibitem[{{Lucy}(1999)}]{Lucy1999}
{Lucy}, L.~B. 1999, \aap, 345, 211

\bibitem[{{Madhusudhan}(2012)}]{Madhusudhan2012}
{Madhusudhan}, N. 2012, \apj, 758, 36

\bibitem[{{Madhusudhan}(2019)}]{Madhusudhan2019}
{Madhusudhan}, N. 2019, \araa, 57, 617

\bibitem[{{Martin} {et~al.}(2014){Martin}, {Nixon}, {Lubow}, {Armitage}, {Price}, {Do{\u{g}}an}, \& {King}}]{Martin2014}
{Martin}, R.~G., {Nixon}, C., {Lubow}, S.~H., {et~al.} 2014, \apjl, 792, L33

\bibitem[{{Marzari} \& {D'Angelo}(2025)}]{Marzari&DAngelo2025}
{Marzari}, F. \& {D'Angelo}, G. 2025, \aap, 695, A53

\bibitem[{{McClure} {et~al.}(2023){McClure}, {Rocha}, {Pontoppidan}, {Crouzet}, {Chu}, {Dartois}, {Lamberts}, {Noble}, {Pendleton}, {Perotti}, {Qasim}, {Rachid}, {Smith}, {Sun}, {Beck}, {Boogert}, {Brown}, {Caselli}, {Charnley}, {Cuppen}, {Dickinson}, {Drozdovskaya}, {Egami}, {Erkal}, {Fraser}, {Garrod}, {Harsono}, {Ioppolo}, {Jim{\'e}nez-Serra}, {Jin}, {J{\o}rgensen}, {Kristensen}, {Lis}, {McCoustra}, {McGuire}, {Melnick}, {{\"O}berg}, {Palumbo}, {Shimonishi}, {Sturm}, {van Dishoeck}, \& {Linnartz}}]{McClure+2023}
{McClure}, M.~K., {Rocha}, W.~R.~M., {Pontoppidan}, K.~M., {et~al.} 2023, Nature Astronomy, 7, 431

\bibitem[{{Minissale} {et~al.}(2022){Minissale}, {Aikawa}, {Bergin}, {Bertin}, {Brown}, {Cazaux}, {Charnley}, {Coutens}, {Cuppen}, {Guzman}, {Linnartz}, {McCoustra}, {Rimola}, {Schrauwen}, {Toubin}, {Ugliengo}, {Watanabe}, {Wakelam}, \& {Dulieu}}]{Minissale+2022}
{Minissale}, M., {Aikawa}, Y., {Bergin}, E., {et~al.} 2022, ACS Earth and Space Chemistry, 6, 597

\bibitem[{{Miranda} \& {Lai}(2015)}]{Miranda+2015}
{Miranda}, R. \& {Lai}, D. 2015, \mnras, 452, 2396

\bibitem[{{Monin} {et~al.}(2007){Monin}, {Clarke}, {Prato}, \& {McCabe}}]{Monin+2007}
{Monin}, J.~L., {Clarke}, C.~J., {Prato}, L., \& {McCabe}, C. 2007, in Protostars and Planets V, ed. B.~{Reipurth}, D.~{Jewitt}, \& K.~{Keil}, 395

\bibitem[{{Muley} {et~al.}(2024){Muley}, {Melon Fuksman}, \& {Klahr}}]{Muley+2024}
{Muley}, D., {Melon Fuksman}, J.~D., \& {Klahr}, H. 2024, \aap, 687, A213

\bibitem[{{Nealon} {et~al.}(2020){Nealon}, {Price}, \& {Pinte}}]{Nealon+2020a}
{Nealon}, R., {Price}, D.~J., \& {Pinte}, C. 2020, \mnras, 493, L143

\bibitem[{{Nelson} \& {Marzari}(2016)}]{Nelson&Marzani2016}
{Nelson}, A.~F. \& {Marzari}, F. 2016, \apj, 827, 93

\bibitem[{{Nixon} {et~al.}(2013){Nixon}, {King}, \& {Price}}]{Nixon+2013}
{Nixon}, C., {King}, A., \& {Price}, D. 2013, \mnras, 434, 1946

\bibitem[{{{\"O}berg} \& {Bergin}(2016)}]{Oberg+2016}
{{\"O}berg}, K.~I. \& {Bergin}, E.~A. 2016, \apjl, 831, L19

\bibitem[{{{\"O}berg} \& {Bergin}(2021)}]{Oberg&Bergin2021}
{{\"O}berg}, K.~I. \& {Bergin}, E.~A. 2021, \physrep, 893, 1

\bibitem[{{{\"O}berg} {et~al.}(2011){{\"O}berg}, {Murray-Clay}, \& {Bergin}}]{Oberg+2011}
{{\"O}berg}, K.~I., {Murray-Clay}, R., \& {Bergin}, E.~A. 2011, \apjl, 743, L16

\bibitem[{{{\"O}berg} \& {Wordsworth}(2019)}]{Oberg&Wordsworth2019}
{{\"O}berg}, K.~I. \& {Wordsworth}, R. 2019, \aj, 158, 194

\bibitem[{{Offner} {et~al.}(2023){Offner}, {Moe}, {Kratter}, {Sadavoy}, {Jensen}, \& {Tobin}}]{Offner+2023}
{Offner}, S.~S.~R., {Moe}, M., {Kratter}, K.~M., {et~al.} 2023, in Astronomical Society of the Pacific Conference Series, Vol. 534, Protostars and Planets VII, ed. S.~{Inutsuka}, Y.~{Aikawa}, T.~{Muto}, K.~{Tomida}, \& M.~{Tamura}, 275

\bibitem[{{Okuzumi} \& {Tazaki}(2019)}]{Okuzumi&Tazaki2019}
{Okuzumi}, S. \& {Tazaki}, R. 2019, \apj, 878, 132

\bibitem[{{Paardekooper} {et~al.}(2023){Paardekooper}, {Dong}, {Duffell}, {Fung}, {Masset}, {Ogilvie}, \& {Tanaka}}]{Paardekooper+2023}
{Paardekooper}, S., {Dong}, R., {Duffell}, P., {et~al.} 2023, in Astronomical Society of the Pacific Conference Series, Vol. 534, Protostars and Planets VII, ed. S.~{Inutsuka}, Y.~{Aikawa}, T.~{Muto}, K.~{Tomida}, \& M.~{Tamura}, 685

\bibitem[{{Paczynski}(1977)}]{Paczynski1977}
{Paczynski}, B. 1977, \apj, 216, 822

\bibitem[{{Pichardo} {et~al.}(2005){Pichardo}, {Sparke}, \& {Aguilar}}]{Pichardo+2005}
{Pichardo}, B., {Sparke}, L.~S., \& {Aguilar}, L.~A. 2005, \mnras, 359, 521

\bibitem[{{Pinte} {et~al.}(2009){Pinte}, {Harries}, {Min}, {Watson}, {Dullemond}, {Woitke}, {M{\'e}nard}, \& {Dur{\'a}n-Rojas}}]{Pinte+2009}
{Pinte}, C., {Harries}, T.~J., {Min}, M., {et~al.} 2009, \aap, 498, 967

\bibitem[{{Pinte} {et~al.}(2006){Pinte}, {M{\'e}nard}, {Duch{\^e}ne}, \& {Bastien}}]{Pinte+2006}
{Pinte}, C., {M{\'e}nard}, F., {Duch{\^e}ne}, G., \& {Bastien}, P. 2006, \aap, 459, 797

\bibitem[{{Pinte} {et~al.}(2019){Pinte}, {van der Plas}, {M{\'e}nard}, {Price}, {Christiaens}, {Hill}, {Mentiplay}, {Ginski}, {Choquet}, {Boehler}, {Duch{\^e}ne}, {Perez}, \& {Casassus}}]{Pinte+2019}
{Pinte}, C., {van der Plas}, G., {M{\'e}nard}, F., {et~al.} 2019, Nature Astronomy, 3, 1109

\bibitem[{{Poblete} {et~al.}(2020){Poblete}, {Calcino}, {Cuello}, {Mac{\'\i}as}, {Ribas}, {Price}, {Cuadra}, \& {Pinte}}]{Poblete+2020}
{Poblete}, P.~P., {Calcino}, J., {Cuello}, N., {et~al.} 2020, \mnras, 496, 2362

\bibitem[{{Poblete} {et~al.}(2019){Poblete}, {Cuello}, \& {Cuadra}}]{Poblete+2019}
{Poblete}, P.~P., {Cuello}, N., \& {Cuadra}, J. 2019, \mnras, 489, 2204

\bibitem[{{Price} \& {Laibe}(2015)}]{Price&Laibe2015}
{Price}, D.~J. \& {Laibe}, G. 2015, \mnras, 451, 813

\bibitem[{{Price} {et~al.}(2018){Price}, {Wurster}, {Tricco}, {Nixon}, {Toupin}, {Pettitt}, {Chan}, {Mentiplay}, {Laibe}, {Glover}, {Dobbs}, {Nealon}, {Liptai}, {Worpel}, {Bonnerot}, {Dipierro}, {Ballabio}, {Ragusa}, {Federrath}, {Iaconi}, {Reichardt}, {Forgan}, {Hutchison}, {Constantino}, {Ayliffe}, {Hirsh}, \& {Lodato}}]{PricePH+2018b}
{Price}, D.~J., {Wurster}, J., {Tricco}, T.~S., {et~al.} 2018, Publications of the Astronomical Society of Australia, 35, e031

\bibitem[{{Pringle}(1981)}]{Pringle1981}
{Pringle}, J.~E. 1981, \araa, 19, 137

\bibitem[{{Raghavan} {et~al.}(2010){Raghavan}, {McAlister}, {Henry}, {Latham}, {Marcy}, {Mason}, {Gies}, {White}, \& {ten Brummelaar}}]{Raghavan+2010}
{Raghavan}, D., {McAlister}, H.~A., {Henry}, T.~J., {et~al.} 2010, \apjs, 190, 1

\bibitem[{{Ragusa} {et~al.}(2017){Ragusa}, {Dipierro}, {Lodato}, {Laibe}, \& {Price}}]{Ragusa+2017}
{Ragusa}, E., {Dipierro}, G., {Lodato}, G., {Laibe}, G., \& {Price}, D.~J. 2017, \mnras, 464, 1449

\bibitem[{{Reipurth} {et~al.}(2014){Reipurth}, {Clarke}, {Boss}, {Goodwin}, {Rodr{\'\i}guez}, {Stassun}, {Tokovinin}, \& {Zinnecker}}]{Reipurth+2014}
{Reipurth}, B., {Clarke}, C.~J., {Boss}, A.~P., {et~al.} 2014, in Protostars and Planets VI, ed. H.~{Beuther}, R.~S. {Klessen}, C.~P. {Dullemond}, \& T.~{Henning}, 267--290

\bibitem[{{Ros} \& {Johansen}(2024)}]{Ros&Johansen2024}
{Ros}, K. \& {Johansen}, A. 2024, \aap, 686, A237

\bibitem[{{Rowther} {et~al.}(2024){Rowther}, {Price}, {Pinte}, {Nealon}, {Meru}, \& {Alexander}}]{Rowther+2024}
{Rowther}, S., {Price}, D.~J., {Pinte}, C., {et~al.} 2024, \mnras, 534, 2277

\bibitem[{{Shakura} \& {Sunyaev}(1973)}]{Shakura&Sunyaev73}
{Shakura}, N.~I. \& {Sunyaev}, R.~A. 1973, \aap, 24, 337

\bibitem[{{Tielens} \& {Allamandola}(1987)}]{Tielens&Allamandola1987}
{Tielens}, A.~G.~G.~M. \& {Allamandola}, L.~J. 1987, in Interstellar Processes, ed. D.~J. {Hollenbach} \& H.~A. {Thronson}, Jr., Vol. 134, 397

\bibitem[{{Tinacci} {et~al.}(2023){Tinacci}, {Germain}, {Pantaleone}, {Ceccarelli}, {Balucani}, \& {Ugliengo}}]{Tinacci+2023}
{Tinacci}, L., {Germain}, A., {Pantaleone}, S., {et~al.} 2023, \apj, 951, 32

\bibitem[{{Tokovinin}(2014)}]{Tokovinin2014}
{Tokovinin}, A. 2014, \aj, 147, 87

\bibitem[{{Tokovinin}(2021)}]{Tokovinin2021}
{Tokovinin}, A. 2021, Universe, 7, 352

\bibitem[{{van't Hoff} {et~al.}(2020){van't Hoff}, {Harsono}, {Tobin}, {Bosman}, {van Dishoeck}, {J{\o}rgensen}, {Miotello}, {Murillo}, \& {Walsh}}]{Hoff+2020}
{van't Hoff}, M. L.~R., {Harsono}, D., {Tobin}, J.~J., {et~al.} 2020, \apj, 901, 166

\bibitem[{{Vorobyov} \& {Basu}(2005)}]{Vorobyov&Basu2005}
{Vorobyov}, E.~I. \& {Basu}, S. 2005, \apjl, 633, L137

\bibitem[{{Vorobyov} {et~al.}(2021){Vorobyov}, {Elbakyan}, {Liu}, \& {Takami}}]{Vorobyov+2021}
{Vorobyov}, E.~I., {Elbakyan}, V.~G., {Liu}, H.~B., \& {Takami}, M. 2021, \aap, 647, A44

\bibitem[{{Zagaria} {et~al.}(2023){Zagaria}, {Rosotti}, {Alexander}, \& {Clarke}}]{Zagaria+2023}
{Zagaria}, F., {Rosotti}, G.~P., {Alexander}, R.~D., \& {Clarke}, C.~J. 2023, European Physical Journal Plus, 138, 25

\bibitem[{{Zurlo} {et~al.}(2024){Zurlo}, {Weber}, {P{\'e}rez}, {Cieza}, {Ginski}, {van Holstein}, {Principe}, {Garufi}, {Hales}, {Kastner}, {Rigliaco}, {Ruane}, {Benisty}, \& {Manara}}]{Zurlo+2024}
{Zurlo}, A., {Weber}, P., {P{\'e}rez}, S., {et~al.} 2024, \aap, 686, A309

\end{thebibliography}

%%--------------------
\appendix 

\section{Molecule freeze-out temperatures}
\label{app:fo}

\begin{table}
\centering
\begin{threeparttable}
\caption{Adopted molecular vibrational frequency, desorption energy, atomic mass, and freeze-out temperature.}
\begin{tabular}{lcccc}
\hline
\hline
% \noalign{\hrule height 1.5pt}
\multicolumn{5}{c}{}\\
\multirow{2}{*}{Molecule} & $\nu_i$\tnote{a} & $E_{i}$\tnote{a} & $m_i$ & $T_{{\rm fo},i}$\tnote{b} \\
 & $[\rm Hz]$ & $[\rm K]$  & $[m_{\rm H}]$ & $[\rm K]$ \\
 \hline
H$_2$O & $4\times10^{13}$ & 5800 & 17.84 & 123 (120)\\
CO$_2$ & $1\times10^{13}$ & 2700 & 43.56 & 58 (50)\\
CO & $7\times10^{11}$ & 1180 & 27.72 & 27 (30)\\
N$_2$ & $8\times10^{11}$ & 1050 & 27.72 & 24 (30)\\
NH$_3$ & $1\times10^{13}$ & 3800 & 16.83 & 83 (80)\\
\hline
\end{tabular}
\begin{tablenotes}
\small
    \item \textbf{Notes. }
    \item[a] Values adopted from table 1 of \citep{Oberg&Wordsworth2019}.
    \item[b] Values computed by assuming $s=0.1\ \micro$m, $d_{\rm gr}=10^{-14}$, and $n_{\rm H}=10^{12}$ cm$^{-3}$. We obtain this last value directly from our simulations. Adopted values are in parentheses. 
\end{tablenotes}
\label{tab:Temp_adp}
\end{threeparttable}
\end{table}

The freeze-out temperature of a molecule is the temperature at which it transitions from a volatile state to ice, adhering to the surface of dust grains. \citet{Hollenbach+2009} offers an approximation for calculating this temperature based on the freeze-out timescale and the thermal desorption rate.
Firstly, the freeze-out timescale relates to the PPD conditions \citep{Tielens&Allamandola1987}
% Pag 435
, including the abundance of dust grains relative to the total number of hydrogen nuclei ($d_{\rm gr}$), the number density of hydrogen ($n_{\rm H}$), gas temperature ($T_{\rm gas}$), the atomic mass of a molecule ($m_i$), and the dust grain size ($s$), expressed as:
\begin{multline}\label{eq:time_fo}    
\tau_{{\rm fo},i} = 9.8\ \left[\mathrm{yr}\right]\times \\ 
    \left(\frac{m_i}{m_{\rm H}}\right)^{0.5} \left(\frac{s}{0.1\ {\rm \micro m}}\right)^{-2} \left(\frac{d_{\rm gr}}{10^{-14}}\right)^{-1} \left(\frac{n_{\rm H}}{10^{10}\ {\rm cm^{-3}}}\right) ^{-1}\left(\frac{T_{\rm gas}}{50\ {\rm K}}\right)^{-0.5}.
\end{multline}
We have assumed a sticking probability of 1.

Finally, the thermal desorption rate relates to the vibrational frequency $(\nu_i)$ and the desorption energy $(E_i)$, with values determined through laboratory experiments, alongside the temperature of the dust grains $(T_{\rm dust})$, expressed as:

\begin{equation}\label{eq:R_fo}    
    R_{{\rm td},i} \simeq \nu_{\rm i} e^{-E_{i}/k_{\rm B}T_{\rm dust}},
\end{equation}
where $k_{\rm B}$ is the Boltzmann’s constant. Note that the temperatures expressed in Equations \ref{eq:time_fo} and \ref{eq:R_fo} are not necessarily the same. We name as the freeze-out temperature $(T_{{\rm fo},i})$ as the temperature at which the condition $R_{{\rm td},i}(T_{{\rm fo},i}) = 1/\tau_{{\rm fo},i}(T_{{\rm fo},i})$ is satisfied. We solve the equation numerically, and Table \ref{tab:Temp_adp} shows the resulting freeze-out temperatures for five molecules. Note that \citet{Ferrero+2020} reported that desorption energy values vary based on molecular structure and ice surface position, resulting in different freeze-out temperatures and consequently shifted snow line positions. Therefore, the values presented in Table \ref{tab:Temp_adp} are estimates of the freeze-out temperatures.

% \begin{figure*}
% \centering
% \begin{center}
%     \includegraphics[width=1.0\textwidth]{Figures/SD_outburst.png} \\
%     % \includegraphics[width=1.0\textwidth]{Figures/SD_outburst.png} 
%     \caption{}
%     \label{fig:SD_out}
% \end{center}
% \end{figure*}

\label{lastpage}
\end{document}